\documentclass[conference]{IEEEtran}
\IEEEoverridecommandlockouts

\usepackage{subfigure}
\usepackage{comment}
\usepackage{booktabs}

\usepackage{cite}
\usepackage{amsmath,amssymb,amsfonts}
\usepackage{algorithmic}
\usepackage{graphicx}
\usepackage{textcomp}
\usepackage{xcolor}
\def\BibTeX{{\rm B\kern-.05em{\sc i\kern-.025em b}\kern-.08em
    T\kern-.1667em\lower.7ex\hbox{E}\kern-.125emX}}

\newcommand{\cpr}[1]{\textcolor{blue}{[cpr: #1]}}

\begin{document}

\title{Design of Robust and Efficient Edge Server Placement and Server Scheduling Policies: Extended  Version}


\author{\IEEEauthorblockN{Shizhen Zhao\thanks{* Shizhen Zhao is the corresponding author.}\IEEEauthorrefmark{2}\IEEEauthorrefmark{1}, Xiao zhang\IEEEauthorrefmark{2}\thanks{\IEEEauthorrefmark{2} Xiao Zhang and Shizhen Zhao are co-first authors.}, Peirui Cao, Xinbing Wang}

\IEEEauthorblockA{\textit{Shanghai Jiao Tong University} \\
\textit{\{shizhenzhao, zhangxiao123, caopeirui, xwang8\}@sjtu.edu.cn}
}
}

\maketitle


\begin{abstract}
We study how to design edge server placement and server scheduling policies under workload uncertainty for 5G networks. We introduce a new metric called \emph{resource pooling factor} to handle unexpected workload bursts. Maximizing this metric offers a strong enhancement on top of robust optimization against workload uncertainty. Using both real traces and synthetic traces, we show that the proposed server placement and server scheduling policies not only demonstrate better robustness against workload uncertainty than existing approaches, but also significantly reduce the cost of service providers. Specifically, in order to achieve close-to-zero workload rejection rate, the proposed server placement policy reduces the number of required edge servers by about 25\% compared with the state-of-the-art approach; the proposed server scheduling policy reduces the energy consumption of edge servers by about 13\% without causing much impact on the service quality.



\end{abstract}





\section{Introduction}
We study the edge server placement problem for 5G networks in this paper. By moving storage, compute, control, etc., closer to the network edge, Edge Computing (EC) could offer higher bandwidth, lower latency and better security to users, and thus has become a key enabling technology for 5G. As 5G takes off, deploying edge computing servers also becomes a priority for service providers.

The key challenge of edge server placement in 5G networks comes from workload uncertainty. 5G adopts \emph{small-cell deployment} to allow end users to communicate at high data rate using millimeter wave. But on the other hand, as the cell size reduces, the number of users served by each base station decreases and thus the aggregated workload at each base station becomes highly variable. One possible solution to handle such workload uncertainty is to over-provision edge computing resources based on the peak workload of all the base stations. However, this approach not only incurs high deployment and energy cost, but also leads to low average resource utilization. 

Most server placement literature~\cite{Wang2019Edge, Guo2019User,Zeng2018Edge,Li2018An,Chin2020Queuing,Cao2020Exploring, Kasi2020Heuristic, Ren2018A, guo2020user, Tero2020Edge, Xu2016Efficient, Jia2017Optimal,Yang2019Cloudlet} does not account for workload uncertainty explicitly. In general, these server placement proposals take a predicted edge workload vector as input, and compute server placement solutions with optimal expenditure, optimal access delay, or minimal energy consumption. However, we cannot keep updating server placement based on the real-time workload. When the real-time workload patterns deviate from the predictions, the performance guarantee offered by these proposals become questionable. Some recent works~\cite{Cui2020Robust, Cui2020Trading, Lu2020Robust, Khan2009Robust} studied how to design edge server placement policies that are robust to server failures. However, workload uncertainty is inherently different from server failures, and thus may need completely new handling mechanisms. 

Handling workload uncertainty is a challenging task. One may use stochastic optimization to handle workload uncertainty. However, this approach requires knowing the detailed distribution of the random workload beforehand, which can be extremely difficult to obtain.
Further, this approach may also suffer from the curse of dimensionality as the edge workload is actually a high-dimensional random vector containing thousands of entries. Another approach to handle uncertainty is robust optimization. This approach formulates uncertainty using a set, and could offer strong performance guarantee as long as the uncertainty is bounded by this set. However, finding an appropriate set for robust optimization can be difficult in practice. If we find a set that only covers a majority, (e.g., 99\%) of the workload patterns, then the robust optimization approach cannot offer any guarantee for the out-of-bound workload patterns. In contrast, if we find a set that covers all the potential workload patterns, this set can be extremely large because the workload uncertainty is heavy-tailed (see Table \ref{tab:statistics} in Section \ref{sec:workload_observation3}), drastically weakening the performance guarantee of robust optimization. Further, in some situations, it may not even be feasible to find such an uncertainty set.

We propose \textbf{RO-RP}, to explicitly account for workload uncertainty in the edge server placement problem. RO-RP is built on top of robust optimization, with newly developed techniques to handle out-of-bound workload patterns. The detailed techniques are described below:
\begin{enumerate}
    \item \textbf{Robust Optimization (RO): } Our trace analysis in Section \ref{sec:workload_observation2} suggests that the edge workload exhibits different patterns during workdays and holidays. Using robust optimization, we can \textbf{optimize server placement based on multiple representative workload patterns}. As a result, we can offer a strong performance guarantee as long as the future workload patterns are within the convex hull formed by these representative workload patterns.
    \item \textbf{Resource Pool Optimization (RP): } However, robust optimization alone cannot offer good guarantee when the future workload patterns are outside of the above convex hull. To overcome this challenge, we introduce a new concept called \textbf{resource pooling factor}. By maximizing this resource pooling factor, the impact of large workload bursts can be minimized.
    \item \textbf{Robust Rounding: } Both of the above two techniques involve solving integer programming problems, which can be computationally expensive. To reduce complexity, we first relax the integer requirement of server placement, and then propose a \textbf{smallest resource pool first} rounding approach to round a fractional solution to an integer solution. This rounding approach turns out to be more robust than other candidate rounding approaches.
\end{enumerate}

In addition to a robust edge server placement solution, we also propose server scheduling to reduce the energy consumption of edge computing. Note that the cumulative edge workload has strong diurnal patterns (see Section \ref{sec:workload_observation1}). Thus, turning off some servers (or changing servers to power-saving mode) during idle hours could potentially save significant amount of energy cost. However, toggling servers between on/off states may incur additional cost. We have explicitly accounted for the switching cost in our server scheduling formulation.

Finally, we evaluate our server placement policy based on both real traces from Shanghai Telecom and synthetic traces. Compared to the existing server placement policies, RO-RP significantly reduces the workload rejection rate given the same number of edge servers. To achieve close-to-zero workload rejection rate, RO-RP requires 25\% fewer edge servers when compared with the state-of-the-art approach. We also evaluate our server scheduling policy using the real trace. Compared to the strategy that turns on all the servers at all times, server scheduling could reduce energy consumption by 13\%.

\section{Related Work}\label{sec:related_work}
Prior work has studied how to deploy edge servers based on workload distribution. However, the workload distribution may not be accurate. Workload uncertainty may have a big impact on the eventual network performance, but unfortunately has not yet received much attention in the existing literature.

Many server placement proposals~\cite{Wang2019Edge, Guo2019User,Zeng2018Edge,Li2018An,Chin2020Queuing,Cao2020Exploring, Kasi2020Heuristic, Ren2018A, guo2020user} have assumed that each base station can be only associated with one edge cloud in their formulations. With this assumption, many standard algorithms, e.g., k-means clustering algorithm, set cover algorithm, etc., can be used to design heuristic solutions for the server placement problem. However, this formulation is inherently non-robust to demand uncertainty. Whenever the workload of a base station bursts, the edge cloud that serves this base station has to bear the burden by itself. In fact, for a given server placement, EC users do benefit from offloading their job requests to multiple edge clouds\cite{farhadi2019service, urgaonkar2015dynamic}. There does exist literature~\cite{Tero2020Edge, Xu2016Efficient, Jia2017Optimal,Yang2019Cloudlet} that allows serving workload from the same base station in different edge clouds. However, workload uncertainty is not considered therein.

To improve the robustness of edge server placement, \cite{Cui2020Robust, Cui2020Trading, Lu2020Robust, Khan2009Robust} studied how to account for server failures in their server placement formulations. However, workload uncertainty is inherently different from server failures, and may happen much more frequently in real time.

One natural idea to deal with workload uncertainty is to use robust optimization. This idea has been used to study the service scheduling problem~\cite{Nguyen2020Two} in EC and the replica server placement problem~\cite{Ho2006Managing} in CDN. However, robust optimization cannot offer any guarantee when the workload is outside of the predicted set. Note that, unlike service scheduling, server placement results cannot be adjusted based on the real-time workload.


In addition to the server installation cost, energy also accounts for a big portion of the cost for edge computing. The energy-optimization literature on edge computing mostly focuses on the edge/IoT devices~\cite{liu2016delay, mao2017survey, lim2018camthings}, but not on the edge servers. In this paper, we study how to perform server on-off scheduling for edge computing to save energy cost. As far as we know, this server scheduling problem was only studied in data centers~\cite{Lin2011Dynamic,goiri2010energy,duy2011prediction,dayarathna2015data}, but has never been explored in edge computing. The biggest difference is that we need to account for the collaboration among different edge clouds when we study server scheduling in edge computing.

\section{Workload Analysis}\label{sec:workload_analysis}

\begin{table*}[!htb]
\centering
\caption{The statistics of workload and inter-arrival time over a time span of six months.}
\label{tab:statistics}
\begin{tabular}{llllllllllllll}
\hline
                   & Max      & Average & std   & 99.999\% & 99.99\%  & 99.9\%  & 99\%   & 90\%   & 80\%  & 70\%  & 60\%  & 50\%  & 40\% \\ \hline
Workload           & 40877    & 2314 & 3283  & 27144   & 10804   & 10800  & 10796 & 8857  & 3910 & 2097 & 1216 & 728  & 499 \\
Inter-arrival time & 13953857 & 7174 & 64180 & 7246866 & 2584387 & 584956 & 82414 & 10731 & 5001 & 3002 & 1990 & 1374 & 899 \\ \hline
\end{tabular}
\end{table*}


The appropriate design of server placement policies requires a deep understanding of the potential workload patterns of EC. However, EC has not seen widespread deployment yet, and thus there may not be any real data for its workloads. Instead, we perform workload analysis based on Shanghai Telecom dataset~\cite{Wang2019Edge,wang2019qos,guo2020user,wang2019delay}, which includes communication records collected about 3042 base stations and 6236620 user requests. We believe that the workload patterns of these communication records can provide a good estimate for the workload patterns of EC.



\subsection{Observation 1: The cumulative workload has strong diurnal patterns}\label{sec:workload_observation1}
\begin{figure*}[!htb]
    \centering
    \subfigure[Number of Requests at different time of the day during one month.] {
        \label{fig:traffic_demands}
        \includegraphics[width=0.64\columnwidth]{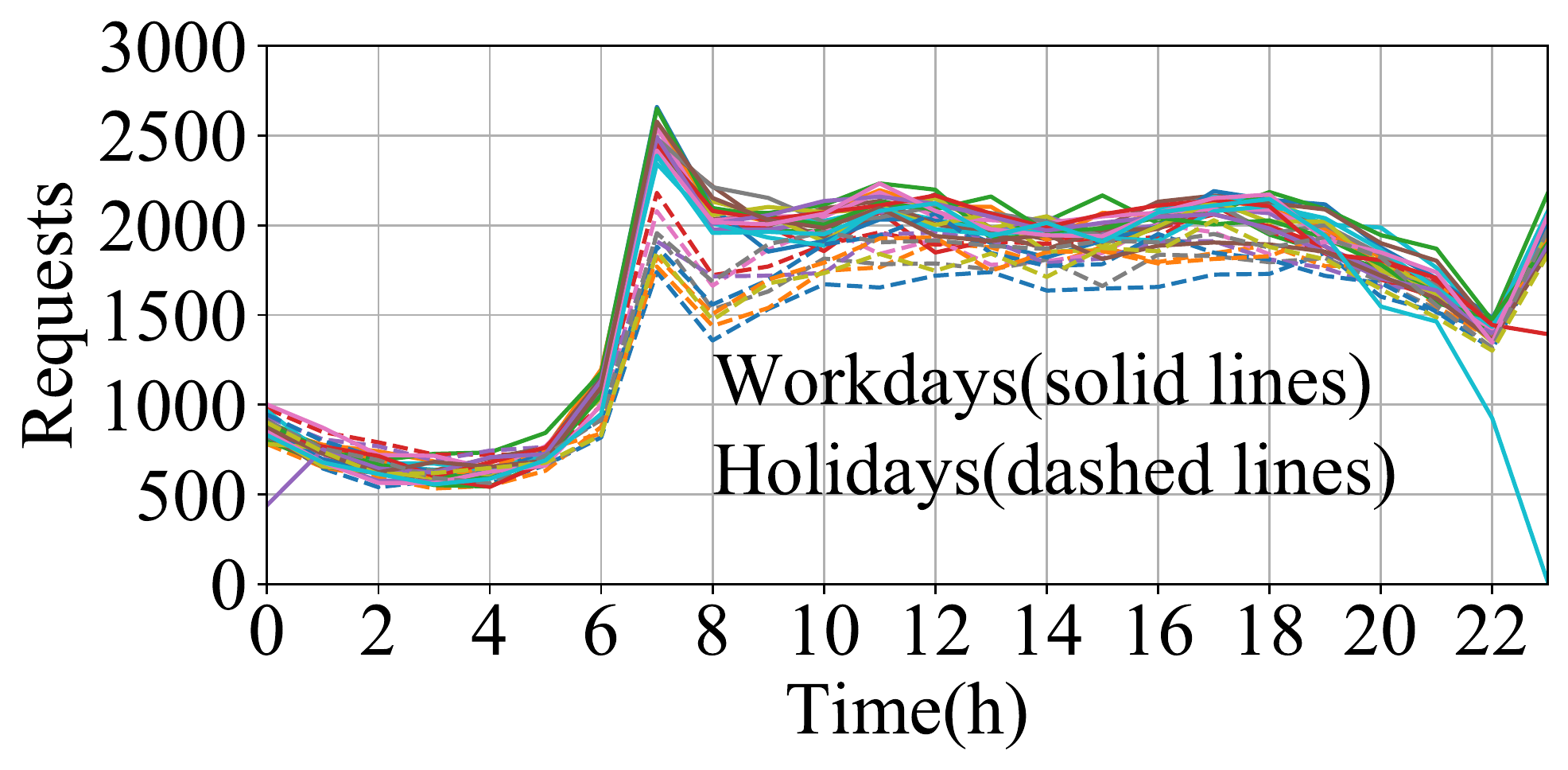}
    }
    \subfigure[Hour-level workload variation for two consecutive weeks.] {
        \label{fig:traffic_delta:a}
        \includegraphics[width=0.64\columnwidth]{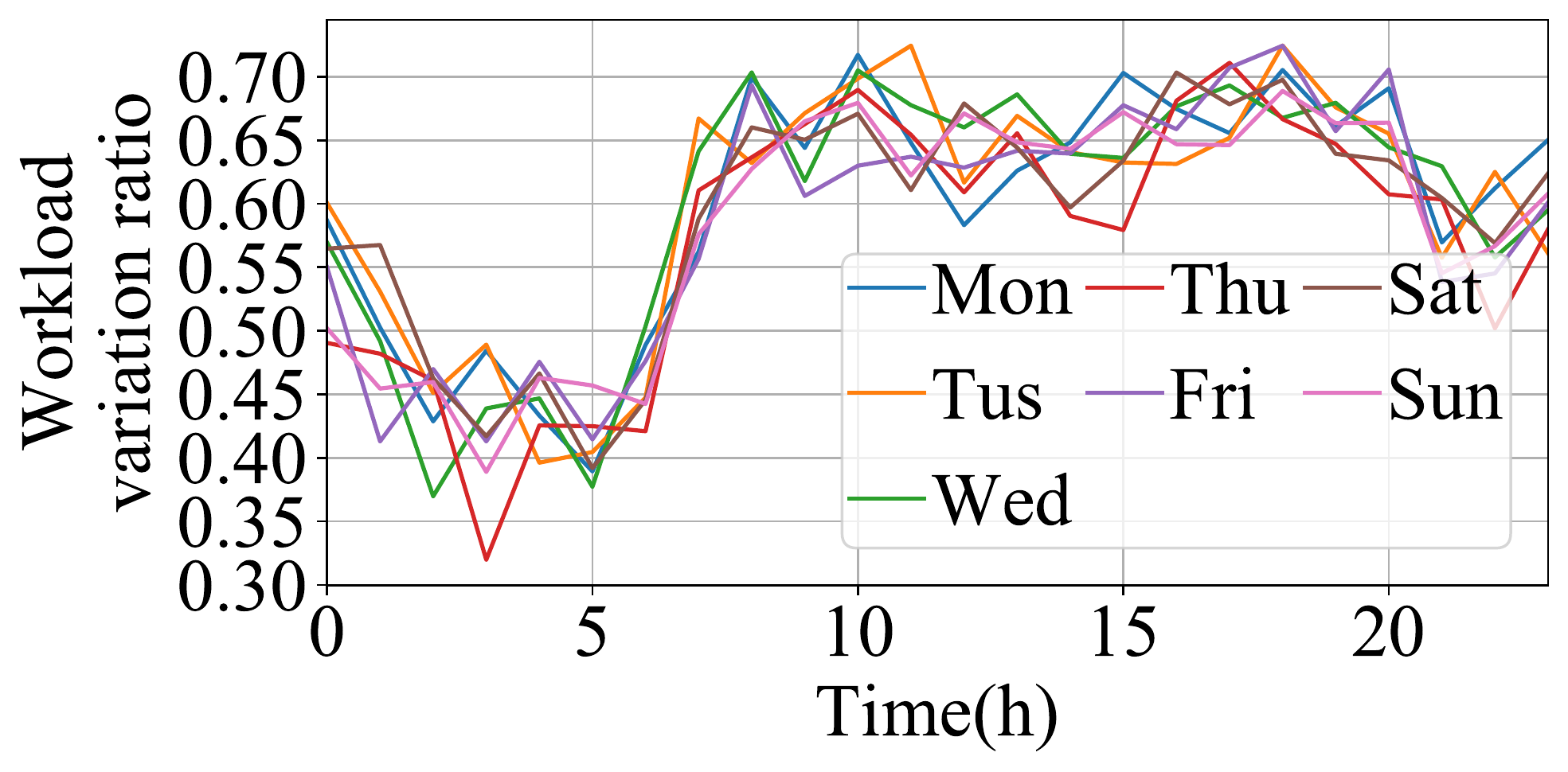}
    }
    \subfigure[Hourly workload statistics for each base station over a period of 6 months.] {
        \label{fig:traffic_delta:b}
        \includegraphics[width=0.64\columnwidth]{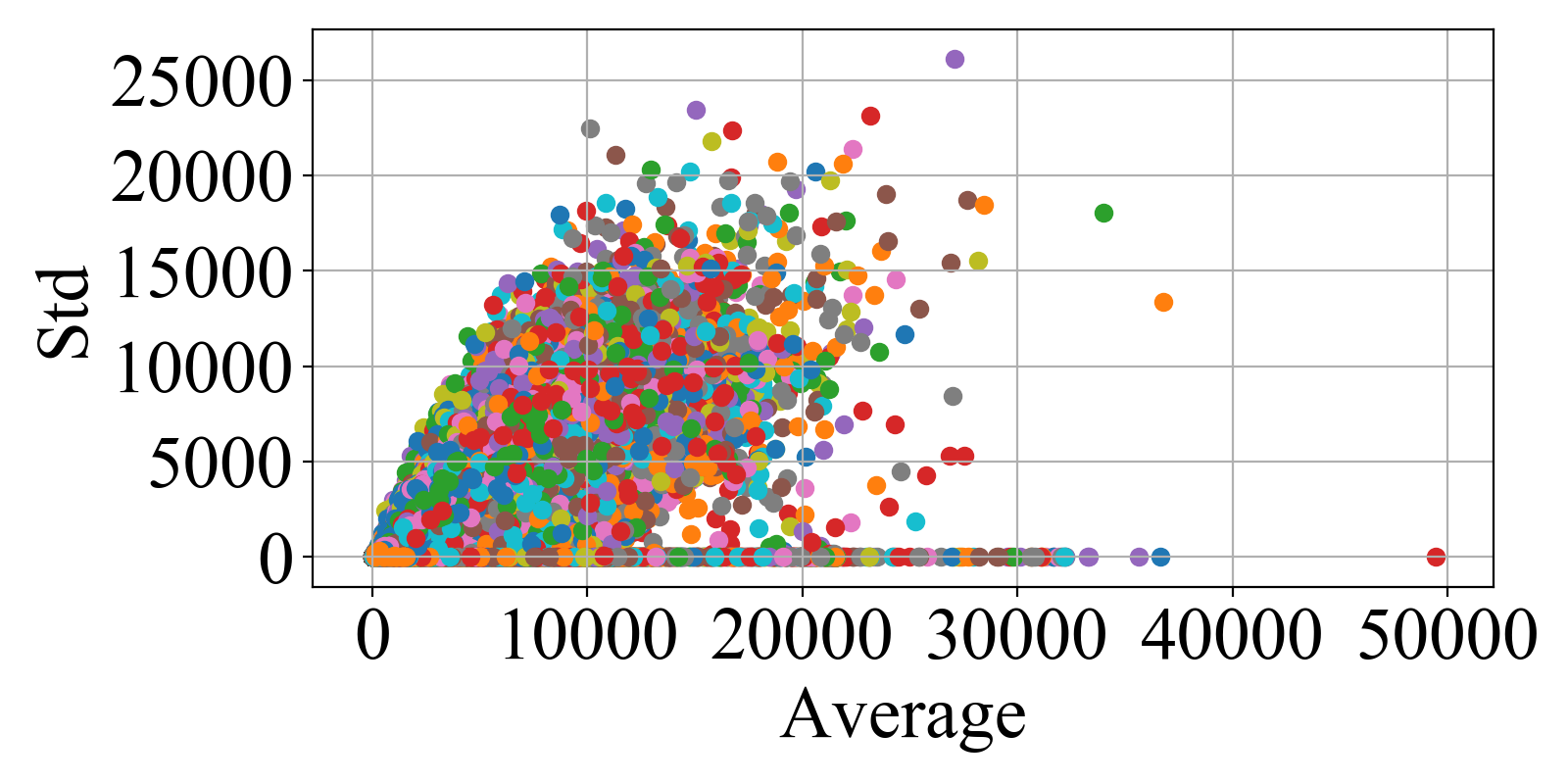}
    }
\caption{Data set analysis.}
\label{fig:traffic_delta}
\end{figure*}

We study how the cumulative workload across all base stations in Shanghai varies at different times of a day using a consecutive of 30 days of data. As shown in Fig.~\ref{fig:traffic_demands}, the total number of requests are the lowest from midnight to about 6:00am every day. As people get up around 6:00-8:00am, the requests rise and peak at around 6:30-7:30am. Then, the requests are relatively consistent until 20:00pm, after which the requests gradually decrease. Another observation from the curves is that the number of requests on workdays (solid lines) is generally higher than that on holidays (dashed lines). This implies that one should not use workday's workload patterns to predict the workload patterns on holidays, and vice versa.


\begin{figure}[!h]
    \centering
    \subfigure[12:00-14:00 on Holiday] {
        \label{fig:workday_weekend0}
        \includegraphics[width=0.8\columnwidth]{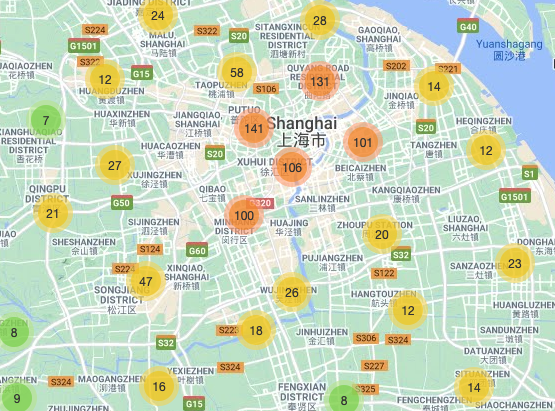}
    }
    \subfigure[12:00-14:00 on Workday] {
        \label{fig:workday_weekend3}
        \includegraphics[width=0.8\columnwidth]{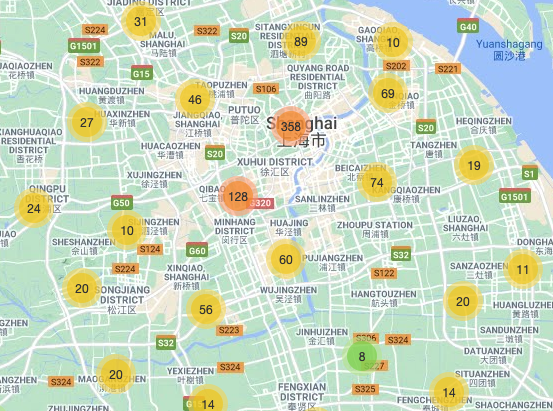}
    }
    \caption{Requests distribution at different days.}
    \label{fig:requests_distribution}
\end{figure}



The strong diurnal patterns of edge workload motivates us to \textbf{perform server on-off scheduling} for EC. The objective is to reduce energy cost, without impacting on service quality.

\subsection{Observation 2: The spatial workload patterns of different days can be highly skewed and dramatically different}\label{sec:workload_observation2}
As shown in Fig.~\ref{fig:requests_distribution}, we randomly pick a holiday and a workday, and study their workload patterns at the same time of a day. Fig.~\ref{fig:traffic_demands} demonstrates that the total workloads at workdays and holidays are approximately the same for most times of a day. However, when we compare the spatial workload distributions in Fig.~\ref{fig:requests_distribution}, the workload patterns are apparently different. This observation suggests that we should \textbf{use multiple different workload patterns to compute server placement solutions}.

Another observation from Fig.~\ref{fig:requests_distribution} is that the workload patterns are highly skewed, with much higher workload in the central area of Shanghai. Hence, traffic agnostic server placement strategies may not perform well, which is verified via simulation in Section \ref{sec:compare_server_placement}.


\subsection{Observation 3: The edge workload is highly bursty}\label{sec:workload_observation3}

To understand the burstiness of edge workload, we analyze the workload sizes and the inter-arrival times of all the requests at each base station over a time span of six months. Table~\ref{tab:statistics} summarizes the average, standard deviation and percentile values of these two metrics. Note that the inter-arrival time is heavy tailed, meaning that it is possible for a base station to receive a large request after being idle for a long period of time.



Since edge workload is highly bursty, accurately predicting edge workload can be very difficult. For example, we can use historical workday/holiday patterns to predict future workload. For the $t$-th hour in every day, we could compute the following workload variation ratio:
$$V_t = \frac{\sum_{m=1}^M{|w_m(t)-w_m^{'}(t)|}}{\sum_{m=1}^M{w_m^{'}(t)}},$$
where $w_m(t)$ is the total workload of the $m$-th base station in the $t$-th hour of the day, and $w_m^{'}(t)$ is the total workload of the $m$-th base station in the $t$-th hour seven days ago. From Fig.~\ref{fig:traffic_delta:a}, we see that the workload variation ratio can be as large as $70\%$. Hence, using historical patterns to predict future workload can be highly inaccurate.


We are also interested in the relationship between the average workload and the workload uncertainty. For every base station and every hour in a week, we collect a sequence of workload values over 6 months with one value per week. We then compute the average value and the standard deviation for every workload sequence, and plot them in Fig.~\ref{fig:traffic_delta:b}. From this figure, we can see that the workload's worst-case standard deviation grows approximately linearly with respect to its average value. However, the range of the standard deviation values is pretty large, and thus there may not be a rule of thumb to accurately predict the workload uncertainty.


The above analysis demonstrates that edge workload is highly variable. Thus, \textbf{how to handle workload uncertainty} becomes a primary focus in this paper.

\section{Model}
\begin{figure}[!htbp]
    \centering
    \subfigure[Edge computing network.] {
        \label{fig:architecture:a}
        \includegraphics[width=0.46\columnwidth]{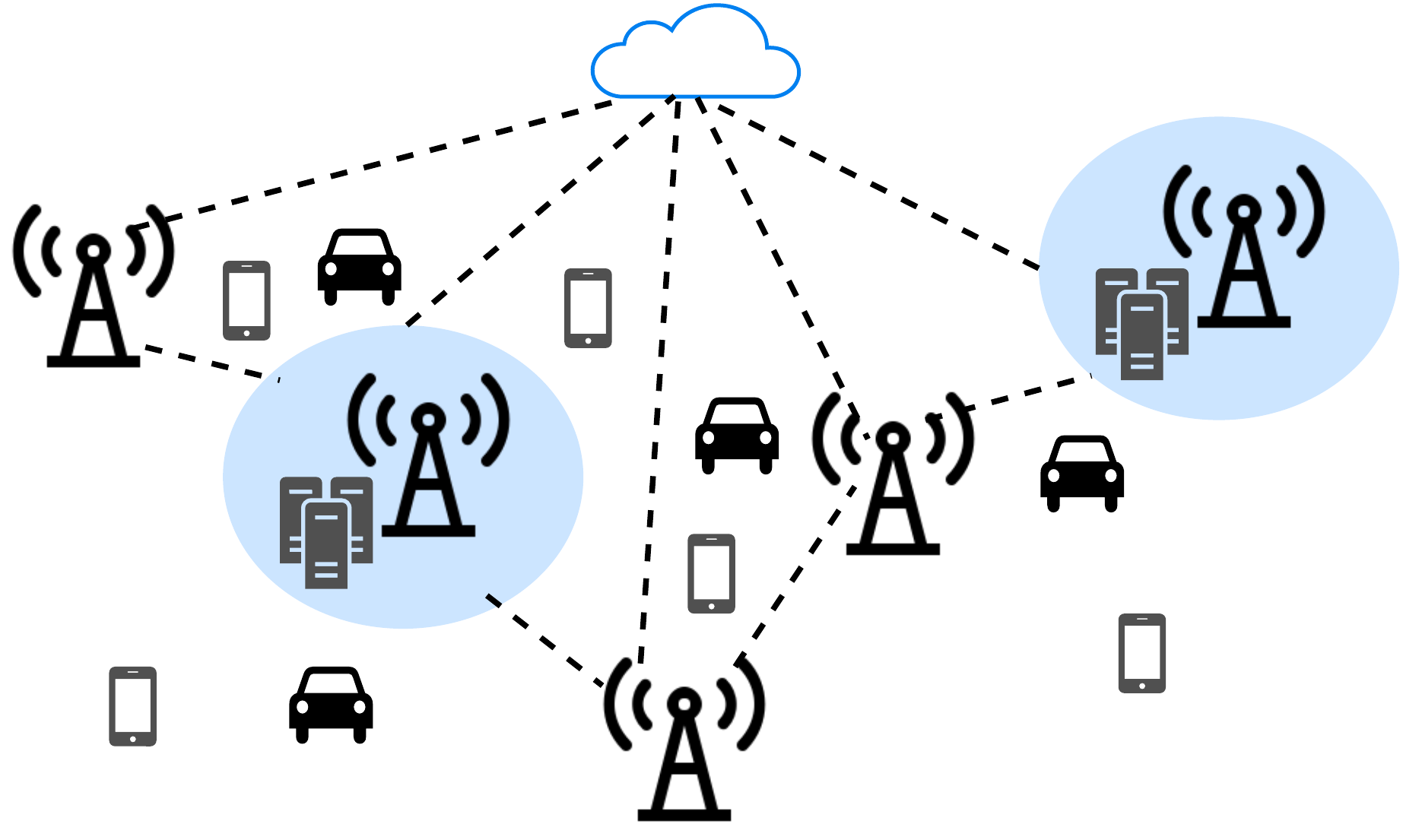}
    }
    \subfigure[Overall workflow.] {
        \label{fig:architecture:b}
        \includegraphics[width=0.46\columnwidth]{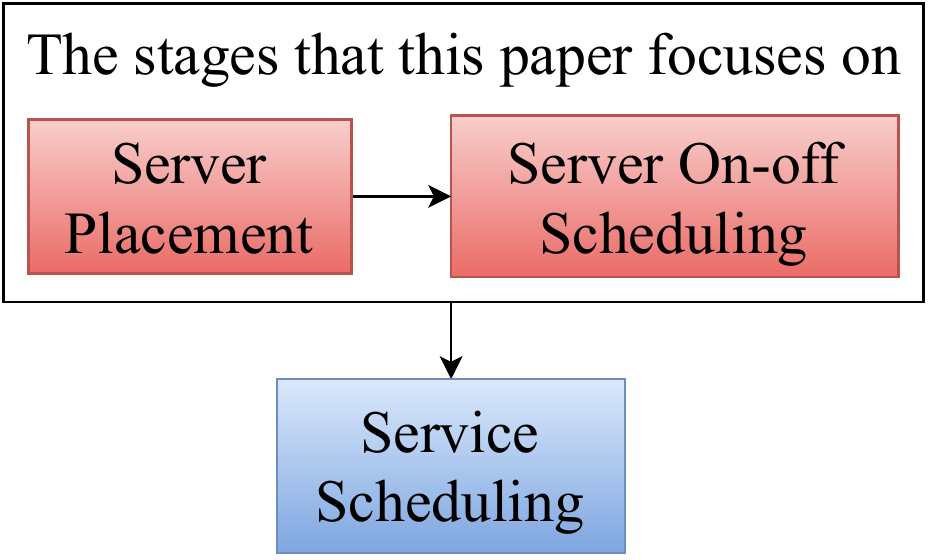}
    }
\caption{Overall Architecture.}
\label{fig:architecture}
\end{figure}

We study edge server placement from a service provider's aspect. A service provider (SP) is responsible for managing the base stations, the central cloud and the edge servers (see Fig. \ref{fig:architecture:a}), with an objective to provide ubiquitous communication and computation services to its users. Users access SP's network through base stations, usually via wireless links. The base stations and the central cloud are typically interconnected through wired links. As we go into the 5G era, when millions of devices connect to the network, and data from each device floods in, edge computing also becomes critical to provide low latency, high reliability, and immense bandwidth.

In general, edge computing consists of two stages: server placement and service scheduling. For server placement, the SP needs to determine where to deploy servers and how many servers to deploy. Typically, edge servers are co-located with base stations. Server placement usually happens at the planning stage. For service scheduling, the SP is responsible for routing users' edge computing requests to nearby edge servers in real time. In this paper, we focus on server placement. Note that running edge servers may incur high energy cost. To reduce energy consumption, we introduce an additional stage, i.e., server on-off scheduling, to make an on-off schedule for edge servers based on predicted workloads. The overall workflow of edge computing is depicted in Fig. \ref{fig:architecture:b}.

\subsection{Mathematical Models}
Let $\{B_1, B_2,...,B_M\}$ denote the set of base stations (BSs) in an 5G network, where $M$ is the total number of base stations. Assume time is slotted. $w_{m}(t)$ denotes the total edge computing workload of BS $B_m$ at slot $t$. The workloads at different base stations form a workload vector $w(t)=[w_{m}(t),m=1,2,...,M]$. Since the edge workload $w_{m}(t)$ has stringent latency requirements, such workload can be only processed by the edge servers deployed at the BS $B_m$ or the nearby BSs of $B_m$. In this paper, we use $\Omega_m$ to denote the set of BSs whose edge servers can serve the workload of the BS $B_m$.

\noindent\textbf{Remark on $\Omega_m$:} One can define $\Omega_m$ using different approaches, including latency requirement, hop-count requirement, distance requirement, $k$-nearest neighbor requirement, etc. The methodology in this paper works for all the possible definitions of $\Omega_m$.

\subsubsection{Server Placement}
Server placement typically happens at the planning stage. Once the edge servers are deployed, it may not be easy to change the server deployment. Hence, server placement must be robust against the potential workload variations. The easiest approach to deal with workload variations is to over-provision edge servers. However, this approach increases both equipment cost and energy cost. In this paper, given a historical trace of workload vectors, we study how to assign a total number of $K$ servers to each BS, such that
\begin{equation}\label{eqn:total_servers}
    S_{1} + S_{2} + \cdots + S_{M} = K,\text{ where $S_m$'s are integers},
\end{equation}
where $S_m$ is the number of edge servers to be deployed at the BS $B_m$.


\subsubsection{Service Scheduling}
Service scheduling happens at the real time stage. At time $t$, for the workload $w_m(t)$ at the BS $B_m$, the objective of service scheduling is to determine the fraction $u_{mn}(t)$ of the workload $w_m(t)$ that is assigned to the base station $B_n\in\Omega_m$. Clearly,
\begin{equation}\label{eqn:demand_split}
\left\{ 
\begin{array}{lc}
\sum_{B_n\in\Omega_m}u_{mn}(t)=1,\text{ for }m=1,2,...,M,\\
0\leq u_{mn}(t)\leq 1\text{ and } u_{mn}(t)=0,\text{ for }B_n\notin \Omega_{m}.
\end{array}
\right. 
\end{equation}

\subsubsection{Server On-off Scheduling}
Server on-off scheduling happens in-between server placement and service scheduling. Fig.~\ref{fig:traffic_demands} suggests that the cumulative edge workload has daily peaks and troughs. Server placement needs to account for the daily peaks. However, keeping all the servers always \emph{on} incurs significant energy cost, especially during idle hours. Let $S(t)=[S_m(t),m=1,2,...,M]$, where $S_m(t)$ is the number of active servers at the BS $B_m$ at time $t$. The objective of server on-off scheduling is to reduce energy cost, without impacting on users' service quality. Clearly,
\begin{equation}\label{eqn:active_servers}
S_m(t)\leq S_m,\forall{m=1,2,...,M}\text{ and }t.
\end{equation}

\subsection{Performance Metrics}
While we design server placement and server on-off scheduling strategies, we are interested in optimizing the following three performance metrics.

\subsubsection{Rejected Workload}
\label{sec:rejected_workload}
When edge servers run out of resources to serve some portion of edge workload, such edge workload is considered as rejected. (Offloading this workload to the central cloud may violate latency requirement.) This could happen when the real-time workload $w(t)=[w_1(t),w_2(t),...,w_M(t)]$ bursts at some BSs. Given the numbers of active servers $S(t)$ and the workload vector $w(t)$, the total amount of rejected workload can be computed by
\begin{equation}\label{formulation:rejected_workload}
\boxed{
\begin{aligned}
\min_{u_{mn}(t)} & \hspace{3mm}\sum_{n=1}^M\max\left\{0,\sum_{m=1}^M w_m(t)u_{mn}(t)-C S_n(t) \right\} &\\
\text{s.t.}& \hspace{3mm} u_{mn}(t)\text{ satisfy (\ref{eqn:demand_split}),}&
\end{aligned}
}
\end{equation}
where $C$ is the capacity of one server. Note that $\max\left\{0,\sum_{m=1}^M w_m(t)u_{mn}(t)-CS_n(t) \right\}$ is the total rejected workload at the BS $B_n$. Solving (\ref{formulation:rejected_workload}) gives the minimum possible amount of workload to be rejected.

\subsubsection{Number of Servers Required}
From service providers' aspect, rejecting EC requests is highly undesirable, because they may lose customer loyalty. Then, another important metrics arises, i.e., what is the minimum number of servers required in order to guarantee zero workload rejection rate? Certainly, this number depends on the server placement strategy and the workload patterns. In Section \ref{sec:compare_server_placement}, we will use extensive simulation to obtain an estimate of this metric for different server placement strategies.

\subsubsection{Cost}
The cost of edge servers consists of two parts:
\begin{itemize}
    \item Server running cost $E_r+E_w\cdot x$, where $E_r$ is the energy cost of running an idle server for one time slot, and $E_w$ is the energy cost per workload unit. Note that $E_r$ may account for more than $50\%$ of the energy cost in a server~\cite{powerconsumption}. 
    \item Switching cost $E_s$, which models the cost of toggling a server between on/off states. As stated in~\cite{Lin2011Dynamic}, if only energy cost matters, then $E_s$ is on the order of the cost of running a server for a few seconds to several minutes; if the increased wear-and-tear is accounted for, then $E_s$ becomes on the order of the cost of running a server for a hour. In this paper, we use the latter to measure the switching cost $E_s$.
\end{itemize}

Note that the total energy cost of all the workload equals $E_w\sum_{m=1}^M\sum_t w_m(t)$, which is out of our control. Hence, in this paper, we mainly focus on the idle-server cost and the switching cost, which in total can be calculated as
\begin{equation}\label{eqn:cost}
    W=\sum_{m=1}^M\sum_t\bigg(E_r\times S_m(t) + E_s\times(S_m(t)-S_m(t-1))^+\bigg),
\end{equation}
where $x^+=\max\{0,x\}$. Clearly, if we choose not to perform server on-off scheduling, then the switching cost becomes zero, but the running cost increases.

\section{Server Placement}\label{sec:server_placement}

The primary objective of server placement is to reduce the workload rejection rate. Since this metric is closely related to service scheduling, our server placement strategy will account for the effect of service scheduling. 

\subsection{A robust joint optimization approach}\label{sec:robust_optimization}
Owing to the fact that edge workload exhibits different patterns at different time of different days, our first idea is to use robust optimization to compute a server placement solution. Specifically, given a sequence of historical workload vectors, we pick multiple representative workload vectors using the following steps:
\begin{description}
\item[Step 1] Divide historical workload vectors into $L$ groups such that the workload vectors in the same group are all 1) from workdays or holidays, and 2) from the same time period (e.g., 8:00-11:59am) of a day.
\item[Step 2] Compute an average workload vector $w^l=[w_m^l,m=1,2,...,M]$ for the $l$-th group of workload vectors\footnote{We have also tried using the component-wise max workload vector to compute a server placement solution. However, our simulation results in Section \ref{sec:average-max-comparison} cannot give any conclusive answer that one option is better than another. Hence, one may use a different approach to compute representative workload vectors.}. Note that the following analysis also works for other choices of workload vectors.
\end{description}
In total, we obtain $L$ workload vectors.

For each workload vector $w^l$, we introduce service scheduling variables $u^l=[u_{mn}^l, m,n=1,2,...,M]$ satisfying (\ref{eqn:demand_split}). Then, we jointly optimize the server placement variables $S=[S_m,m=1,2,...,M]$ satisfying (\ref{eqn:total_servers}) and the service scheduling variables $[u^1,u^2,...,u^L]$. 

For each workload vector $w^l$ and the corresponding service scheduling variables $u^l$, it is easy to obtain the total workload allocated to the BS $B_n$, i.e., $\sum_{m=1}^M w_m^l u_{mn}^l$. To reduce the likelihood of workload rejection, we impose the following constraint that restricts the edge server utilization to be less than $\beta$ for any BS $B_n$ and any workload vector $w^l$:
\begin{equation}\label{eqn:server_capacity}
\sum_{m=1}^M w_m^l u_{mn}^l\leq S_n C \beta,\text{ for any }B_n\text{ and }w^l.
\end{equation}
Then, the overall formulation is given as
\begin{equation}\label{formulation:server_placement}
\boxed{
\begin{aligned}
\min_{S,u^l,\beta} & \hspace{20mm}\beta &\\
\text{s.t.}& \hspace{3mm} S\text{ satisfy (\ref{eqn:total_servers}),}&\\
& \hspace{3mm} S,u^l,\beta\text{ satisfy (\ref{eqn:demand_split})(\ref{eqn:server_capacity}) for }l=1,2,...,L.&
\end{aligned}
}
\end{equation}

\noindent\textbf{Understanding the performance guarantee of (\ref{formulation:server_placement}):} Let $\hat{S},\hat{u}^l,\beta^*$ be the optimal solution of (\ref{formulation:server_placement}). Consider a future workload vector $w\leq\sum_{l=1}^L\lambda_l w^l$, where $\lambda_l\geq 0, l=1,2,...,L$ and $\sum_{l=1}^L \lambda_l=1$. In other words, the workload vector $w$ is bounded by the convex hull of the workload vectors $[w^1,w^2,...,w^L]$. We construct the service scheduling variable $u$ for $w$ as follows:
\begin{equation}\label{eqn:calculateu}
u_{mn}=\sum_{l=1}^L\frac{\lambda_l w_m^l}{\sum_{i=1}^L \lambda_i w_m^i}\hat{u}_{mn}^l
\end{equation}
It is easy to check that $u$ satisfies (\ref{eqn:demand_split}). Further,
\begin{eqnarray}
\sum_{m=1}^M w_m u_{mn}&=&\sum_{m=1}^M w_m \sum_{l=1}^L\frac{\lambda_l w_m^l}{\sum_{i=1}^L \lambda_i w_m^i}\hat{u}_{mn}^l\nonumber\\
&\leq&\sum_{m=1}^M\sum_{l=1}^L\lambda_l w_m^l\hat{u}_{mn}^l=\sum_{l=1}^L\lambda_l\sum_{m=1}^M w_m^l\hat{u}_{mn}^l\nonumber\\
&\leq&\sum_{l=1}^L\lambda_l \hat{S}_n C\beta^*=\hat{S}_n C\beta^*.\nonumber
\end{eqnarray}
Hence, the workload $w$ is supportable by the server placement solution $\hat{S}$ with max server utilization no greater than $\beta^*$.

\noindent\textbf{Understanding the drawback of (\ref{formulation:server_placement}):} 
When future workload vectors are bounded by the convex hull of $[w^1,w^2,...,w^L]$, the above analysis indicates that (\ref{formulation:server_placement}) can offer strong performance guarantee. However, some future workload vectors can be out of bound, owing to the fact that edge workload is highly bursty. Certainly, one can increase the convex hull by scaling up the representative workload vectors $[w^1,w^2,...,w^L]$. Unfortunately, $\beta^*$ will also increase, making the performance guarantee of robust optimization weaker. Further, if $\beta^*>1$, then the performance guarantee becomes useless.

\subsection{Handling Out-of-bound Workload with Resource Pooling}
\label{sec:resource_pooling}
\begin{figure}[!htb]
    \centering
    \subfigure[Small resource pool.] {
        \label{fig:motivation:a}
        \includegraphics[width=0.46\columnwidth]{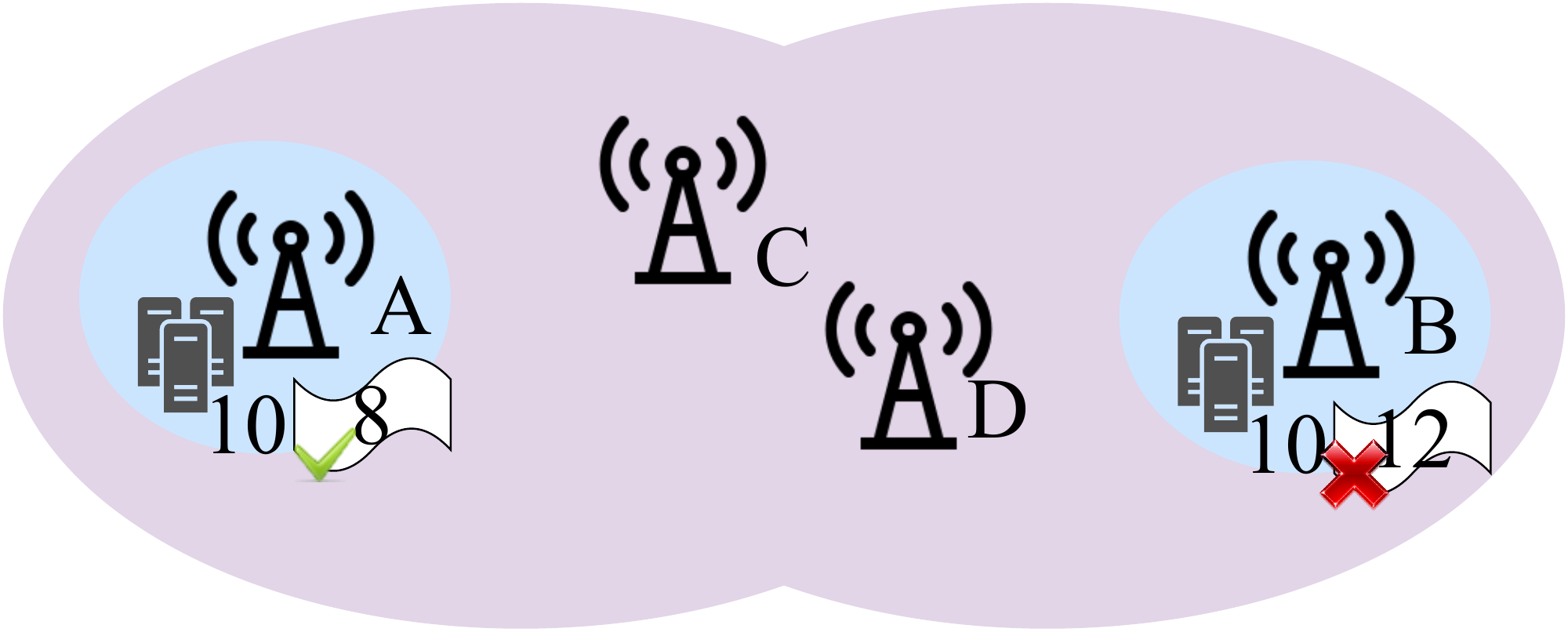}
    }
    \subfigure[Large resource pool.] {
        \label{fig:motivation:b}
        \includegraphics[width=0.46\columnwidth]{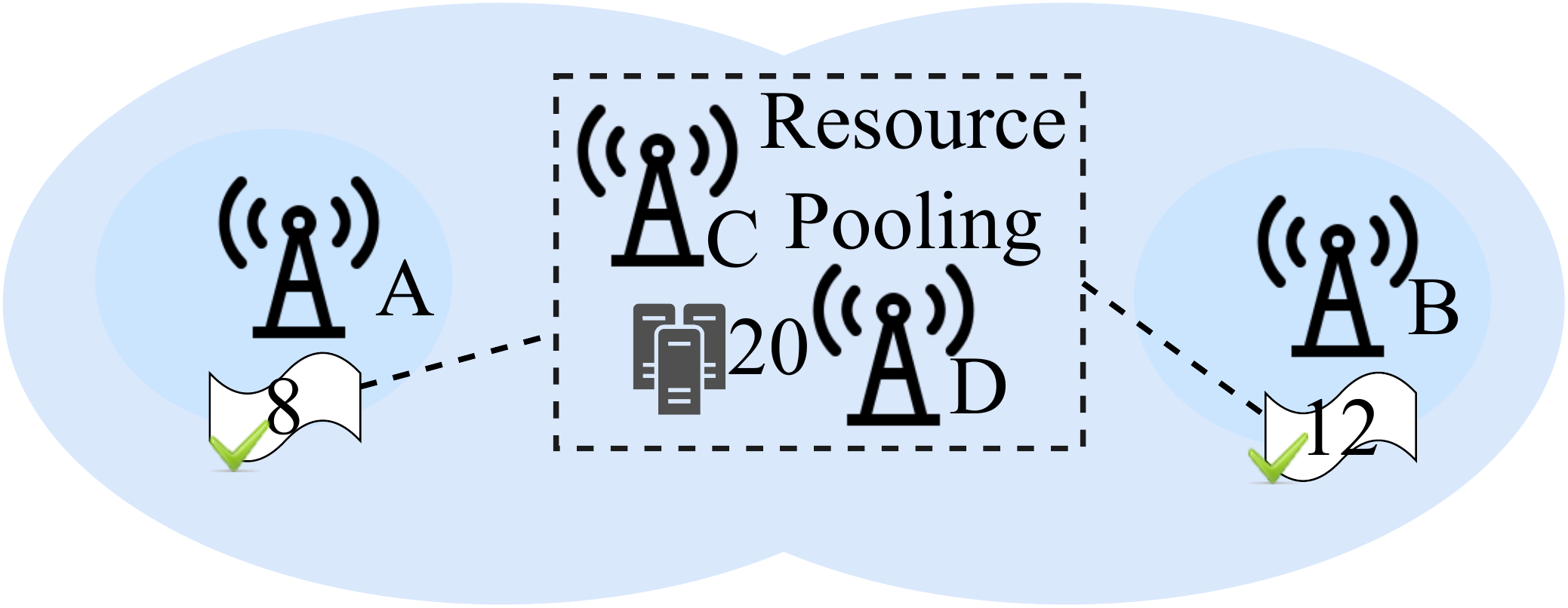}
    }
\caption{Motivation example of resource pooling.}
\label{fig:motivation}
\end{figure}

\noindent\textbf{Intuition:} We first introduce the concept of \emph{resource pool}. For the workload at the BS $B_m$, its \emph{resource pool} is defined as the total amount of server resources that can serve the workload, which equals to $\sum_{B_n\in \Omega_m}S_n$. To understand how resource pooling helps mitigate the impact of out-of-bound workload, we consider the motivation example in Fig. \ref{fig:motivation}. The predicted MEC workloads at the base stations $A$ and $B$ are both $10$, while the real time workloads turn out to be $8$ and $12$. If we deploy $10$ units of computing resources at $A$ and $B$ each (see Fig. \ref{fig:motivation:a}), then in real time, $2$ units of $B$'s workload will be rejected owing to the fact that 1) $B$'s MEC servers are already fully utilized, and thus cannot serve $B$'s burst; 2) $A$ is too far from $B$, and thus cannot serve $B$'s burst either. On the other hand, if we deploy computing resources at the base stations $C$ and $D$ (see Fig. \ref{fig:motivation:b}), the computing resources of $C$ and $D$ can serve the workloads from both $A$ and $B$, because $C$ and $D$ are both within an acceptable distance from $A$ and $B$. As a result, the resource pools of both $A$ and $B$ increases from $10$ to $20$. Then, as $B$'s workload bursts from $10$ to $12$, we can offload $2$ units of workload from the edge cloud $D$ to the edge cloud $C$. With an increase resource pool, the workload from the base station $B$ will not be rejected any more. 

\noindent\textbf{Formulation:} Motivated by the above example, we introduce a resource pooling factor $\eta$, such that
\begin{equation}\label{eqn:resource_pool_factor}
    \eta \max_{l=1}^L \{w_m^l\} \leq \sum_{B_n\in \Omega_m}S_n.
\end{equation}
Then, we compute the server placement result by maximizing the resource pooling factor $\eta$:
\begin{equation}\label{formulation:server_placement_robustification}
\boxed{
\begin{aligned}
&\hspace{-5mm}\textbf{Maximize resource pooling factor:}&\\
\max_{S,u^l,\eta} & \hspace{20mm} \eta &\\
\text{s.t.}& \hspace{3mm} S,\eta \text{ satisfy (\ref{eqn:total_servers})(\ref{eqn:resource_pool_factor}),}&\\
& \hspace{3mm} S,u^l,\beta^*\text{ satisfy (\ref{eqn:demand_split})(\ref{eqn:server_capacity}) for }l=1,2,...,L.&
\end{aligned}
}
\end{equation}
Note that our server placement policy involves two steps: 1) compute $\beta^*$ using the \emph{robust optimization} formulation (\ref{formulation:server_placement}); 2) optimize \emph{resource pooling} based on (\ref{formulation:server_placement_robustification}). Hence, we will also use \textbf{RO-RP} to represent our server placement policy.

\noindent\textbf{Understanding the performance guarantee of (\ref{formulation:server_placement_robustification}):}
Let $S^*, \eta^*$ be the optimal solution of (\ref{formulation:server_placement_robustification}). 
Consider an arbitrary workload vector $w$. If the workload vector $w$ is within the convex hull of $[w^1,w^2,...,w^L]$, then the max server utilization can be bounded by $\beta^{*}$. 
Otherwise, we could find a $w^{'}$ in the convex hull, such that the distance between $w$ and $w^{'}$ is minimized. We can view $w$ as a workload vector bursted on top of $w^{'}$.
Assume that the workload at a base station $B_m$ increases by a percentage of $\gamma$, i.e., $w_m=w_m^{'}(1+\gamma)$.
Then, if we load balance the additional workload $w_m^{'} \gamma$ uniformly among all the edge servers at base stations $B_n\in\Omega_m$ (other workloads are load balanced based on the service scheduling ratio computed by (\ref{eqn:calculateu})), then the maximum server utilization will be upper bounded by
$\beta^*+\frac{w_m^{'} \gamma}{\sum_{n\in \Omega_m}S^*_m}\leq \beta^*+\frac{\gamma}{\eta^*}$. Certainly, if we only solve (\ref{formulation:server_placement}) to obtain a server placement solution, we can still compute a resource pooling factor. However, this resource pooling factor can be much smaller than the optimal value $\eta^*$. As a result, the maximum server utilization can be much higher in case of workload burst. When the maximum server utilization is above one, some workload would be rejected.


\subsection{Reduce Algorithmic Complexity for Server Placement}
\label{sec:reducec_algorithmic_complexity}
We compute server placement results by solving (\ref{formulation:server_placement}) and (\ref{formulation:server_placement_robustification}). However, both (\ref{formulation:server_placement}) and (\ref{formulation:server_placement_robustification}) are integer programming problems, which are computationally expensive. In fact, we have tried solving (\ref{formulation:server_placement_robustification}) directly based on the Shanghai Telecom dataset with 3042 base stations, but unfortunately the state-of-the-art integer programming solver, Gurobi~\cite{gurobi}, cannot finish with a solution even after running a few hours.

To reduce computational complexity, we adopt a \emph{relaxing and rounding} approach. Specifically, we first allow the server placement variables $S$ to take fractional values. Then, both (\ref{formulation:server_placement})\footnote{The constraints (\ref{eqn:server_capacity}) contains a multiplicity term $S_n\beta$. To convert (\ref{formulation:server_placement}) into a linear programming problem, we need to substitute $S_n$ by $\hat{S}_n=S_n\beta$ in (\ref{eqn:server_capacity}), and replace (\ref{eqn:total_servers}) by $\sum_{m=1}^M \hat{S}_n = K\beta$ in (\ref{formulation:server_placement}).} and (\ref{formulation:server_placement_robustification}) can be converted to linear programming problems, which can be solved in polynomial time. After obtaining a fractional server placement solution of $S$, we can then round the fractional solution to an integer solution. Note that we need to ensure that (\ref{eqn:total_servers}) is satisfied after rounding.

Let $\tilde{S}^*=[\tilde{S}_1^*,\tilde{S}_2^*,...,\tilde{S}_M^*]$ be a fractional server placement solution. We have tried different rounding schemes as follows:
\begin{enumerate}
    \item \textbf{Smallest Resource Pool First (SRPF):} Each base station $B_m$ has a resource pool, whose size is $\sum_{B_n\in\Omega_m}\tilde{S}_n^*$. The base stations with smaller resource pools are given higher priority to round up its fractional solution.
    \item \textbf{Random Rounding (RR):} Randomly select some base stations to round up their fractional server counts.
    \item \textbf{Largest Resource Pool First (LRPF):} Contrary to the Smallest Resource Pool First policy, the base stations with larger resource pools are given higher priority to round up its fractional solution.
    \item \textbf{Largest Decimal First (LDF)}: The base stations with larger decimal part ($\tilde{S}_m^*-\lfloor \tilde{S}_m^*\rfloor$) are given higher priority to round up their fractional server counts.
    \item \textbf{Largest Scale-down First (LSF)}: When we round down a number $\tilde{S}_m^*$, the server placement solution is scaled down by $\frac{\tilde{S}_m^*-\lfloor \tilde{S}_m^*\rfloor}{\tilde{S}_m^*}$. The base stations with larger scale-down values are given higher priority to round up.
\end{enumerate}

Eventually, We adopt the Smallest Resource Pool First approach to perform rounding. This approach is the most robust against workload uncertainty, and our simulation results in Section \ref{sec:evaluate_rounding} further show that this approach yields the lowest workload rejection rate under bursty workload.

\section{Server Scheduling}
When service providers deploy edge servers, they must ensure that there are enough servers even during peak hours. However, edge workloads exhibit strong diurnal pattern (see Figure~\ref{fig:traffic_demands}). Then, during idle hours, these edge servers may incur significant energy consumption unnecessarily, thus increasing the cost of edge computing.

To reduce cost, we study how to dynamically adjust the number of active servers based on the time-varying workload. When workload is low, we turn off some servers (or change to power saving mode) to save energy; when workload becomes higher, we will turn on some servers. 

We perform server on-off scheduling on daily basis. Specifically, We divide a day into $N$ equal-length time intervals $T=\{t_1,t_2,...,t_N\}$, where $t_n$ is the timestamp of the middle point of the $n$-th interval. At the day-ahead planning stage, we first predict the average workload vector $\bar{w}(t_n)=[\bar{w}_1(t_n), \bar{w}_2(t_n),...,\bar{w}_M(t_n)]$ in each time interval $t_n$ based on historical traces, and then compute a server on-off schedule for the following day.

\subsection{Server Scheduling without Considering Switching Cost}
We first ignore the switching cost of toggling servers between on/off states. Then, we can study server scheduling in each time interval separately. In this case, the server scheduling problem becomes very similar to the server placement problem, and thus the methodology used in Section \ref{sec:server_placement} can be also applied here. The only difference is that we only need to account for one workload vector in server scheduling.

Let $K(t)$ be the total number of active servers at time $t$. Then, the number of active servers $S(t)=[S_1(t),S_2(t),...,S_M(t)]$ at time $t$ must satisfy (\ref{eqn:active_servers}) as well as the following constraint:
\begin{equation}\label{eqn:scheduling_server_num}
    \sum_{m=1}^M S_{m}(t) = K(t).
\end{equation}
Let $\beta(t)$ be the maximum server utilization at time $t$ and $u(t) = [u_{mn}(t),m,n = 1,2,...,M]$ be the service scheduling variables satisfying (\ref{eqn:demand_split}). Then, we can convert  (\ref{eqn:server_capacity}) to
\begin{equation}\label{eqn:scheduling_capacity}
\sum_{m=1}^M \bar{w}_m(t) u_{mn}(t)\leq S_n(t) C \beta(t).
\end{equation}
Let $\eta(t)$ be the resource pooling factor at time $t$. Then, the constraint (\ref{eqn:resource_pool_factor}) can be rewritten as
\begin{equation}\label{eqn:scheduling_resource_pool_factor}
    \eta(t) \bar{w}_m \leq \sum_{B_n\in \Omega_m}S_n(t).
\end{equation}

Then, we can compute a server scheduling solution at time $t$ in two steps:

\noindent\textbf{Step 1:} Compute the optimal $\beta^*(t)$ by solving
\begin{equation}\label{formulation:server_scheduling_no_switching_cost_step1}
\boxed{
\begin{aligned}
\min_{S(t),u(t),\beta(t)} & \hspace{20mm}\beta(t) &\\
\text{s.t.}\hspace{5mm}& S(t),u(t),\beta(t)\text{ satisfy (\ref{eqn:demand_split})(\ref{eqn:active_servers})(\ref{eqn:scheduling_server_num})(\ref{eqn:scheduling_capacity}).}&
\end{aligned}
}
\end{equation}
\noindent\textbf{Step 2:} Compute the optimal $S^*(t)$ and $\eta^*(t)$ by solving
\begin{equation}\label{formulation:server_scheduling_no_switching_cost_step2}
\boxed{
\begin{aligned}
\max_{S(t),u(t),\eta(t)} & \hspace{20mm}\eta(t) &\\
\text{s.t.}\hspace{6mm}& \hspace{-3mm}S(t),u(t),\beta^*(t),\eta(t)\text{ satisfy (\ref{eqn:demand_split})(\ref{eqn:active_servers})(\ref{eqn:scheduling_server_num})(\ref{eqn:scheduling_capacity})(\ref{eqn:scheduling_resource_pool_factor}).}&
\end{aligned}
}
\end{equation}

\begin{figure}[!htb]
    \centering
    \includegraphics[width=0.8\linewidth]{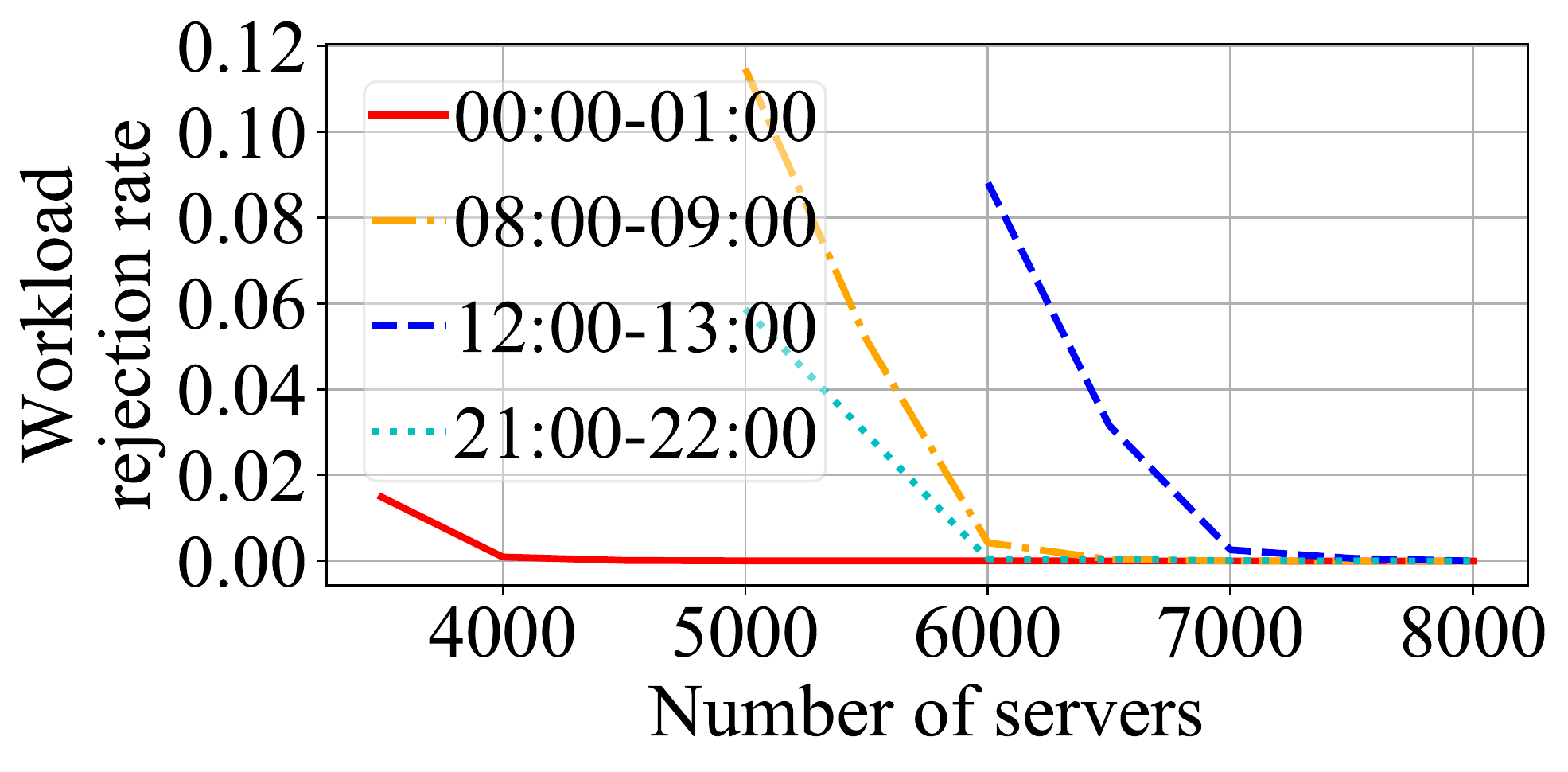}
    \caption{Workload rejection rate under different number of servers in different periods.}
    \label{fig:draw_server_scheduling_error_workload2}
\end{figure}

Clearly, for different $K(t)$, we can obtain different solutions of $S^*(t)$. To understand how many active servers are required, we use historical traces to calculate the fraction of rejected workload. From Fig.~\ref{fig:draw_server_scheduling_error_workload2}, we can see that different number of active servers are required to ensure low workload rejection rate. Specifically, the workload between time 0:00 and 1:00 is low, and thus approximately 4500 servers are sufficient to offer close-to-zero workload rejection rate. In contrast, the workload at 08:00-09:00, 12:00-13:00 and 21:00-22:00 is much higher, thus requiring more active servers.

Based on the above analysis, we can pick a $K^*(t)$ that leads to close-to-zero workload rejection rate for each time interval $t$, and then use $K^*(t)$ to compute $S^*(t)$.




\subsection{Considering Switching Cost for Server Scheduling}
We account for the switching cost for server scheduling in this section. To ensure close-to-zero workload rejection rate, we slightly modify the constraint (\ref{eqn:scheduling_server_num}) as
\begin{equation}\label{eqn:scheduling_server_num_guarantee}
    \sum_{m=1}^M S_{m}(t) \geq K^*(t).
\end{equation}
When $K(t)=K^*(t)$, we can solve (\ref{formulation:server_scheduling_no_switching_cost_step1}) and (\ref{formulation:server_scheduling_no_switching_cost_step2}) to obtain $\beta^*(t)$ and $\eta^*(t)$ first. Then, we can modify (\ref{eqn:scheduling_capacity}) and (\ref{eqn:scheduling_resource_pool_factor}) as
\begin{equation}\label{eqn:scheduling_capacity_guarantee}
\sum_{m=1}^M \bar{w}_m(t) u_{mn}(t)\leq S_n(t) C \beta^*(t).
\end{equation}
\begin{equation}\label{eqn:scheduling_resource_pool_factor_guarantee}
    \eta^*(t) \bar{w}_m \leq \sum_{B_n\in \Omega_m}S_n(t).
\end{equation}

Then, we can compute a server scheduling solution by minimizing the cost function (\ref{eqn:cost}) as follows:
\begin{equation}\label{formulation:server_scheduling}
\boxed{
\begin{aligned}
&\hspace{-5mm}\textbf{Server On-off Scheduling:}&\\
\min_{S(t)} & \hspace{2mm}\sum_{m=1}^M\sum_t\bigg(E_r S_m(t) + E_s(S_m(t)-S_m(t-1))^+\bigg) &\\
\text{s.t.}\hspace{1mm}& \hspace{2mm}S(t),u(t)\text{ satisfy (\ref{eqn:demand_split})(\ref{eqn:active_servers})(\ref{eqn:scheduling_server_num_guarantee})(\ref{eqn:scheduling_capacity_guarantee})(\ref{eqn:scheduling_resource_pool_factor_guarantee}) for any }t.&
\end{aligned}
}
\end{equation}
Note that the term ``$E_s(S_m(t)-S_m(t-1))^+$'' captures the switching cost.

\noindent\textbf{Remark: } We need to solve (\ref{formulation:server_scheduling_no_switching_cost_step1}) and (\ref{formulation:server_scheduling_no_switching_cost_step2}) multiple times to obtain $K^*(t),\beta^*(t)$ and $\eta^*(t)$ before solving (\ref{formulation:server_scheduling}). This pre-solving step is critical to ensure low workload rejection rate.

\noindent\textbf{Remark on Algorithmic Complexity:} Although server on-off scheduling is only calculated once everyday, directly solving (\ref{formulation:server_scheduling_no_switching_cost_step1}), (\ref{formulation:server_scheduling_no_switching_cost_step2}) and (\ref{formulation:server_scheduling}) is still computationally prohibitive. To reduce complexity, we can adopt the same \emph{relaxing and rounding} approach proposed in Section \ref{sec:reducec_algorithmic_complexity}.

\section{Evaluation}
\subsection{Introduce the Trace for Evaluation}
\subsubsection{Shanghai Telecom's Real Communication Records}
We have introduced this real trace in Section \ref{sec:workload_analysis}. This trace contains communication records for 3042 base stations and 6236620 user requests over a time span of six months. We will use this trace to compare our server placement and server scheduling solution against previous approaches. 


\subsubsection{Synthetic Trace}
It was shown in Section \ref{sec:workload_observation3} that edge workload is hard to predict accurately. Hence, our solution must be robust against potential workload bursts. Unfortunately, the historical real traces may not offer a comprehensive coverage over the burst patterns. Hence, we construct synthetic traces to evaluate solution robustness.

Our synthetic traces are constructed as follows. Given an arbitrary real workload vector $w=[w_1,w_2,...,w_M]$, we randomly select a set  containing 100-200 BSs out of the 3024 BSs, choose a scaling factor in $\{1.2, 1.5, 1.8, 2\}$ and then scale up the workload of all the selected BSs by this scaling factor. For the same real workload vector, we repeat the above step multiple times and generate a total of 240 bursty workload vectors. Note that we generate workload bursts by multiplying the original workload by a scaling factor. The reason is that the workload with the highest variability has its standard deviation scaling approximately linearly with respect to its average value (see Fig. ~\ref{fig:traffic_delta:b} in Section \ref{sec:workload_analysis}).



\subsection{Compare Different Server Placement Policies}\label{sec:compare_server_placement}
We group existing server placement policies into three categories and evaluate them using the real trace.

\noindent\textbf{Traffic-agnostic policies}: This policy does not use any workload information for server placement. We evaluate the following policies in this category:
\begin{enumerate}
\item \textbf{Random}: Place $K$ servers at randomly chosen BSs.

\item \textbf{Clustering}: Use k-means algorithm to group BSs into $k$ clusters. Let $s_i^{(C)}$ be the number of BSs in the $i$-th cluster. Place $K\frac{s_i^{(C)}}{\sum_{i=1}^k s_i^{(C)}}$ servers at the centroid of the $i$-th cluster.

\item \textbf{Uniform}: Divide the geographical area into fixed-sized zones. Let $s_i^{(U)}$ be the number of BSs in the $i$-th zone. Place $K\frac{s_i^{(U)}}{\sum_{i=1}^k s_i^{(U)}}$ servers at the centroid of the $i$-th zone.
\end{enumerate}

\noindent\textbf{Traffic-aware but uncertainty-agnostic policies}: This policy computes server placement based on a predicted workload vector, but does not account for the workload uncertainty. We evaluate two policies in this category:
\begin{enumerate}
\item \textbf{Traffic-aware without load balancing (TwithoutLB)}: In this policy, the workload at each BS can be only allocated to the nearest BS with edge servers. Many existing server placement policies fall into this category~\cite{Wang2019Edge, Guo2019User,Zeng2018Edge,Li2018An,Chin2020Queuing,Cao2020Exploring, Kasi2020Heuristic, Ren2018A, guo2020user}. In this paper, we use k-means algorithm to group BSs into $k$ clusters, and assign all the workload in each cluster to the centroid of this cluster. Let $l_i$ be the total workload of the BSs in the $i$-th cluster. Place $K\frac{l_i}{\sum_{i=1}^k l_i}$ servers at the centroid of the $i$-th cluster.

\item \textbf{Traffic-aware with load balancing (TwithLB)}: In this policy, the workload at each BS can be load balanced to the close-by BSs. This setting was also adopted in \cite{Tero2020Edge, Xu2016Efficient, Jia2017Optimal,Yang2019Cloudlet}. Here, we compute a server placement solution by solving (\ref{formulation:server_placement}) with only one workload vector. (We use historical average workload vector in the evaluation.)
\end{enumerate}

\noindent \textbf{Traffic and uncertainty-aware policies}: This is our server placement strategy proposed in Section \ref{sec:server_placement} (namely RO-RP).










\begin{figure*}[!htb]
    \centering
    \subfigure[Workload rejection rate for six server placement policies.] {
        \label{fig:draw_history_compare_error}
        \includegraphics[width=0.64\columnwidth]{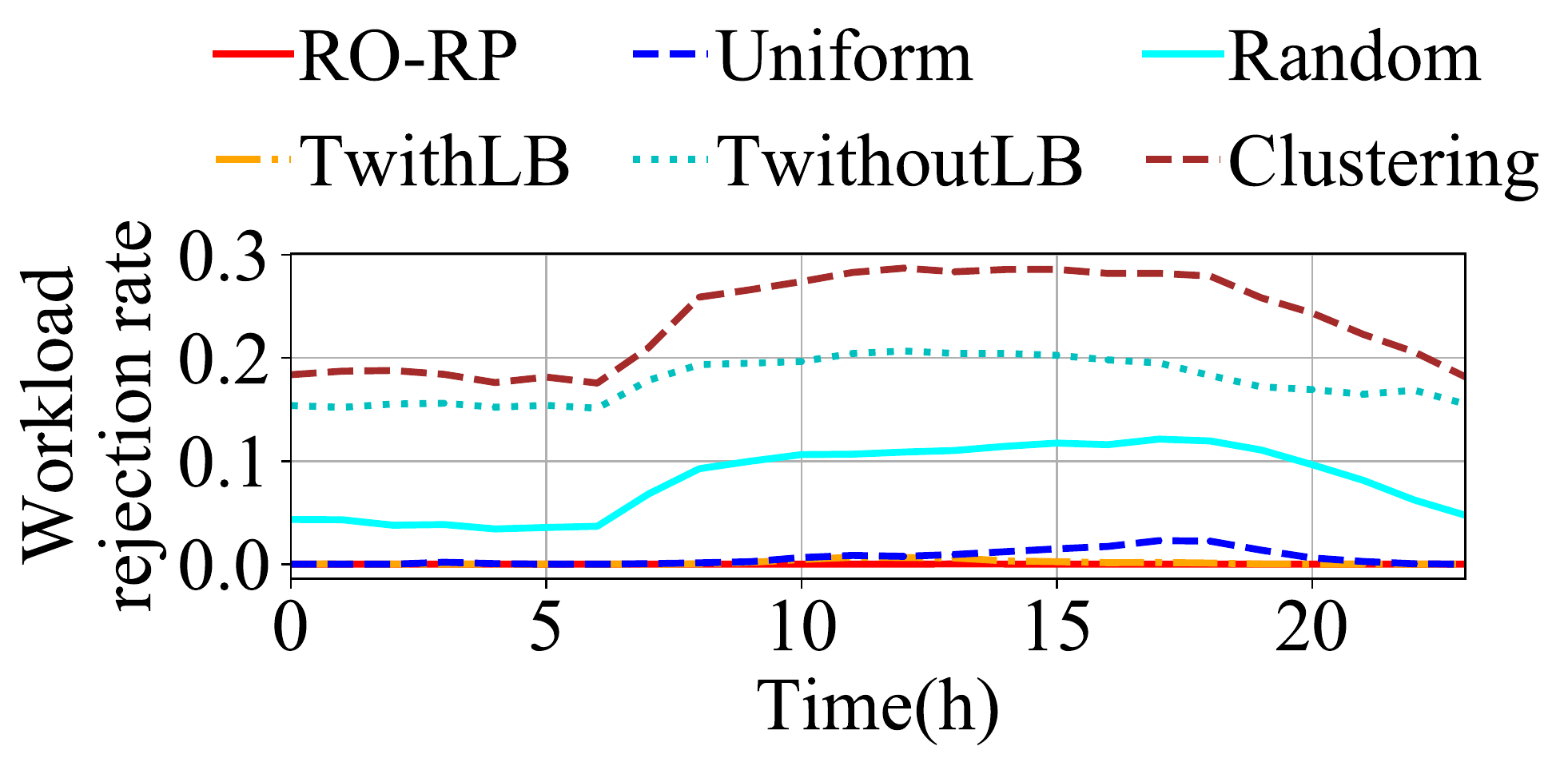}
    }
    \subfigure[A zoom in of the performance for the top-3 best server placement policies.] {
        \label{fig:draw_history_compare_error_b}
        \includegraphics[width=0.64\columnwidth]{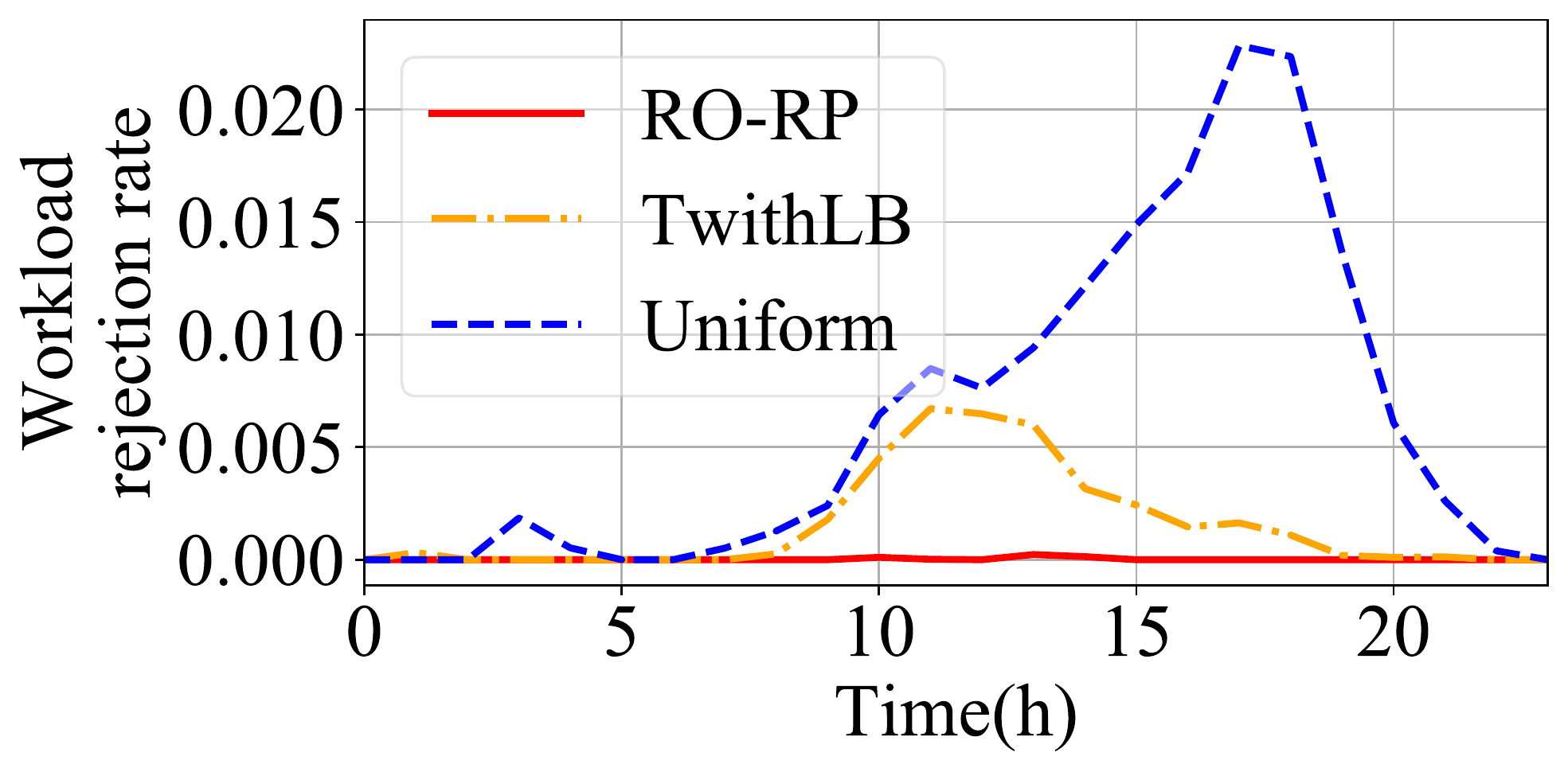}
    }
    \subfigure[Number of servers required for the top-3 best server placement policies.] {
        \label{fig:draw_error_server_plot}
        \includegraphics[width=0.64\columnwidth]{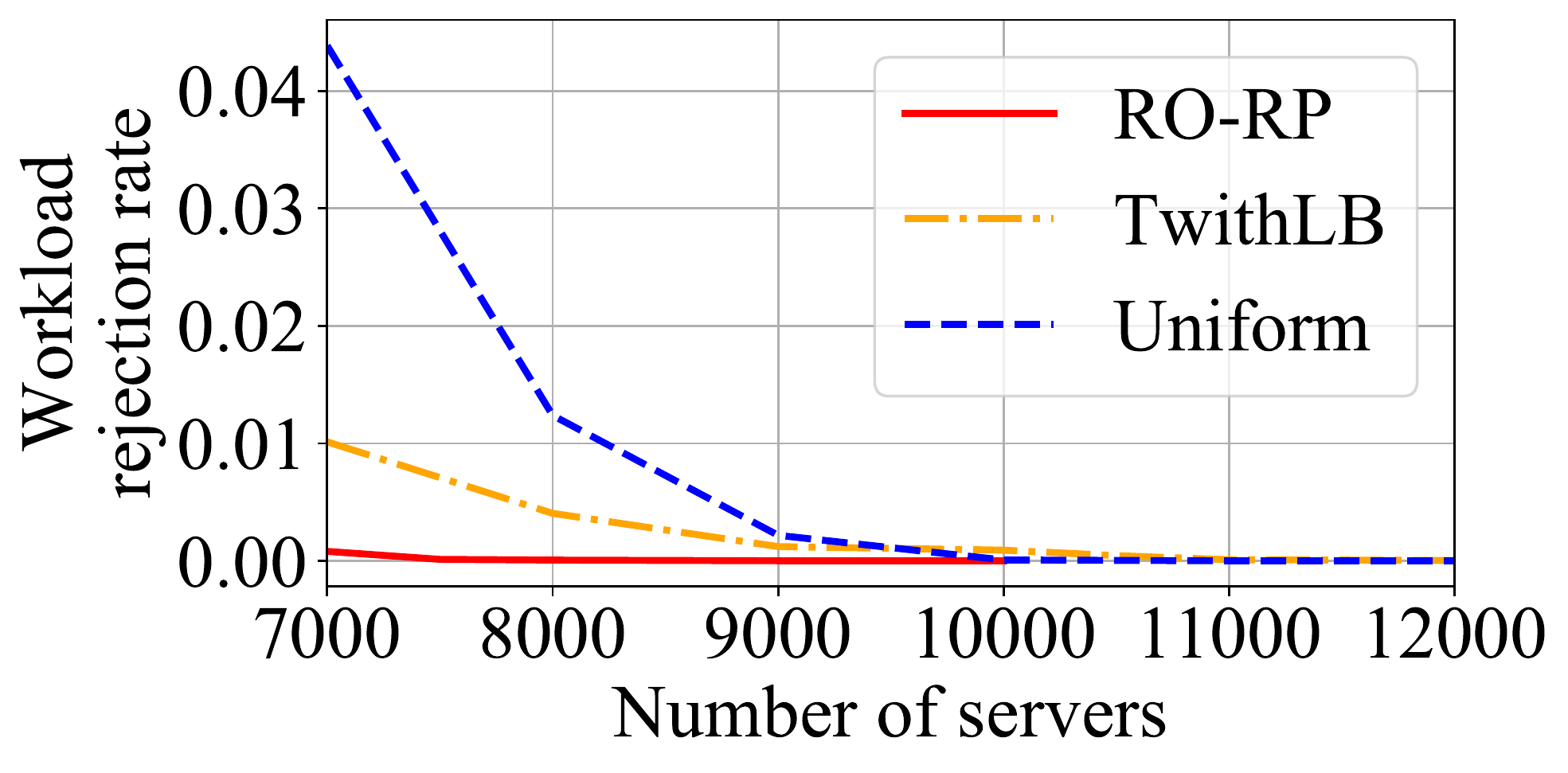}
    }
\caption{Comparison of different server placement policies.}
\label{fig:diff_method_error}
\end{figure*}

\begin{figure*}[!htb]
    \centering
    \subfigure[Performance of different rounding schemes.] {
        \label{fig:overload_round_error_rate_cdf}
        \includegraphics[width=0.64\columnwidth]{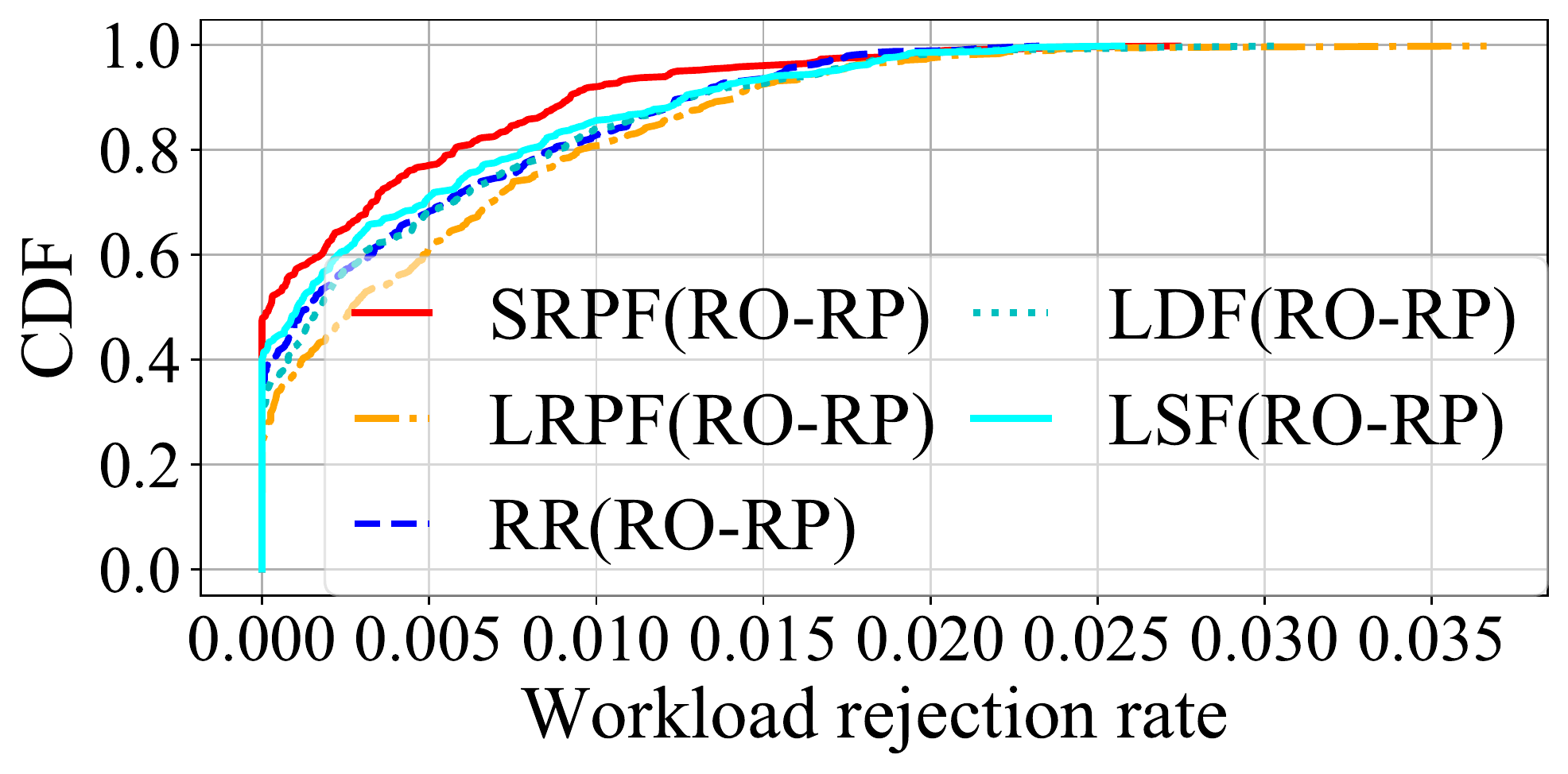}
    }
    \subfigure[Average workload vs. component-wise max workload.] {
        \label{fig:overload_robust_balance_pool}
        \includegraphics[width=0.64\columnwidth]{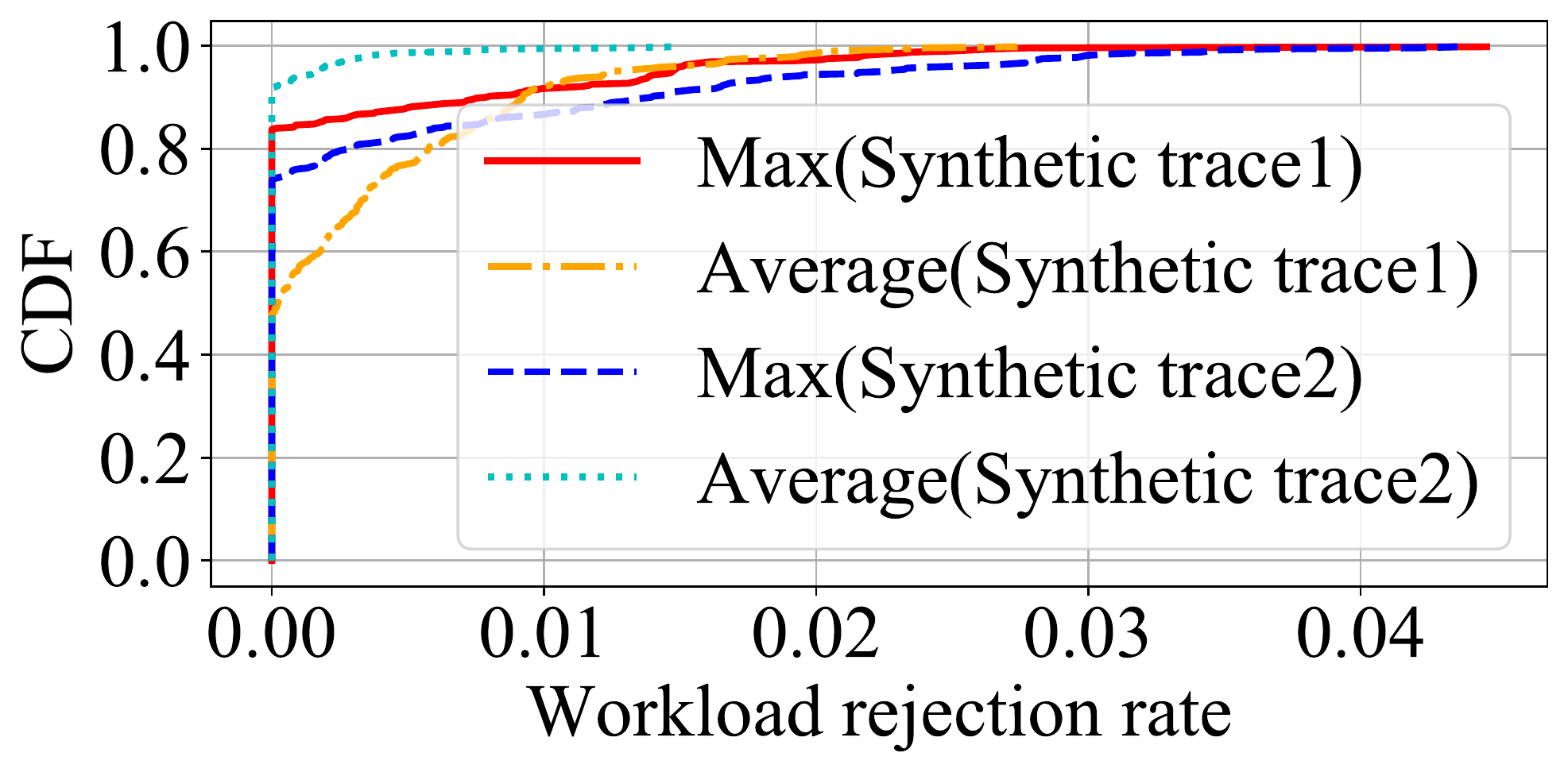}
    }
    \subfigure[Contribution of robust optimization and resource pooling.] {
        \label{fig:overload_error_rate_cdf_80000}
        \includegraphics[width=0.64\columnwidth]{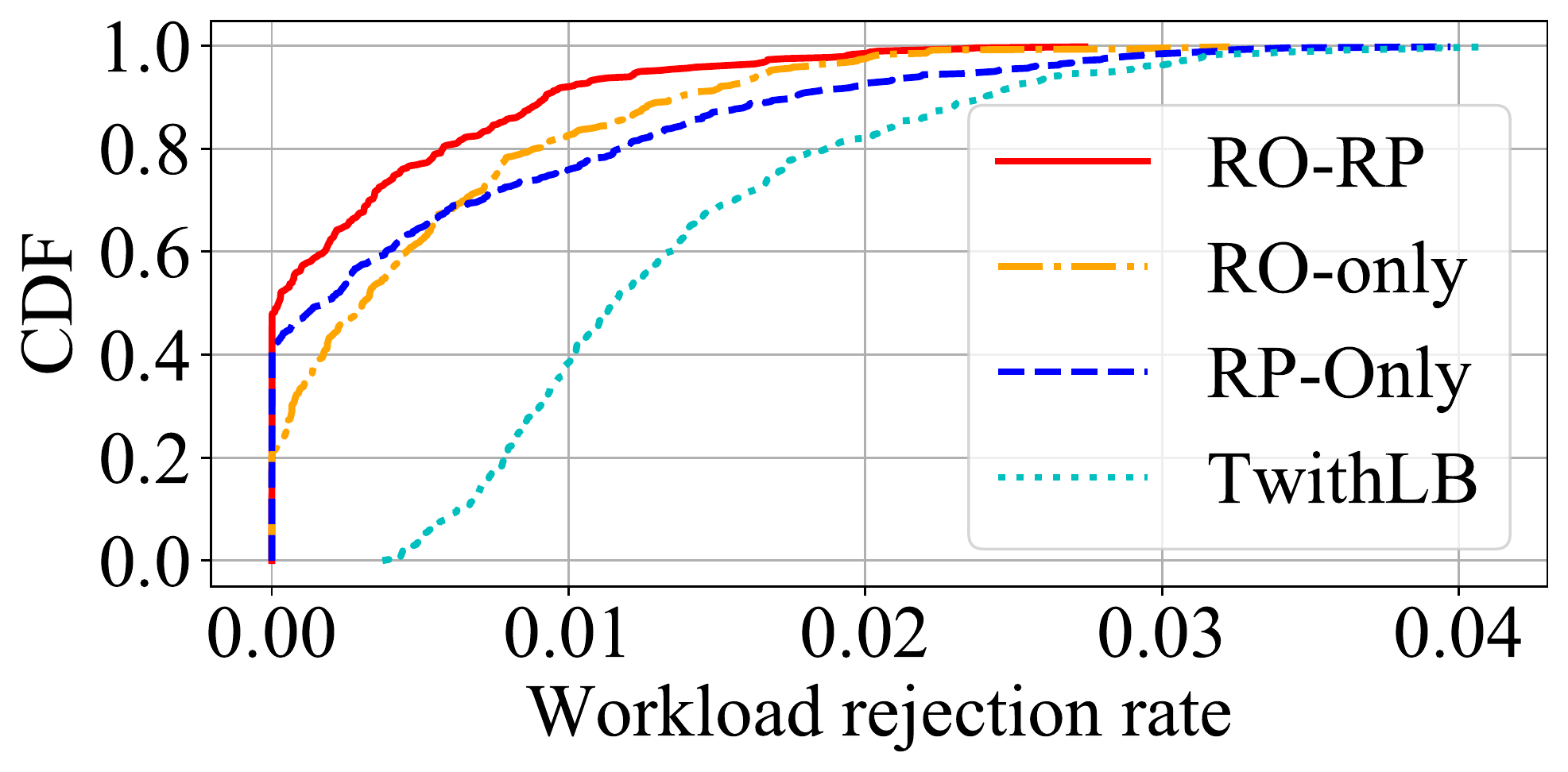}
    }
\caption{Different design choices of our server placement policy.}
\end{figure*}

\begin{figure*}[!htb]
    \centering
    \subfigure[The switching cost is equal to running a server in idle state for one hour.] {
        \label{fig:draw_energy_consumption_1}
        \includegraphics[width=0.64\columnwidth]{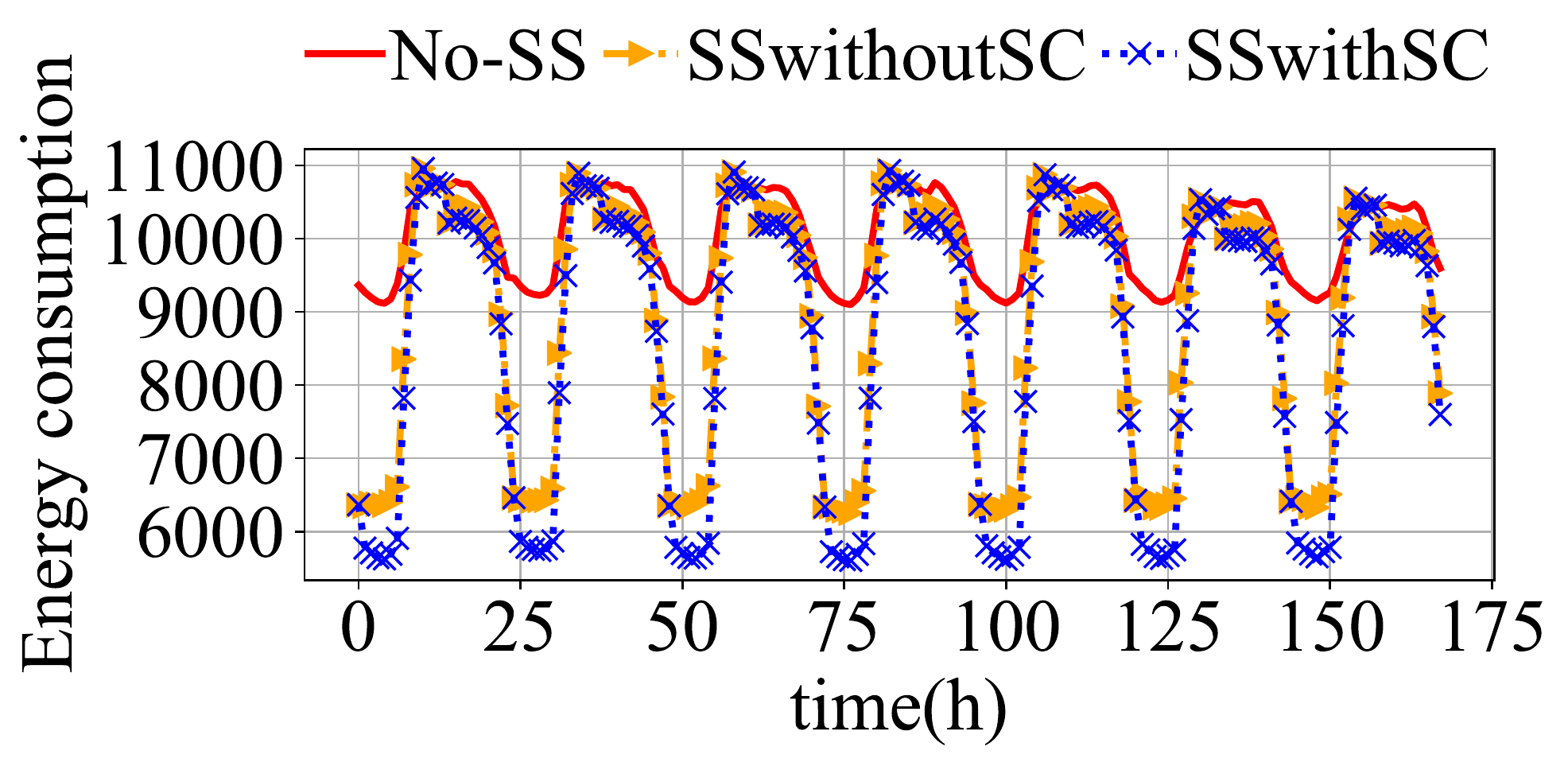}
    }
    \subfigure[The switching cost is equal to running a server in idle state for 12 minutes.] {
        \label{fig:draw_energy_consumption_02}
        \includegraphics[width=0.64\columnwidth]{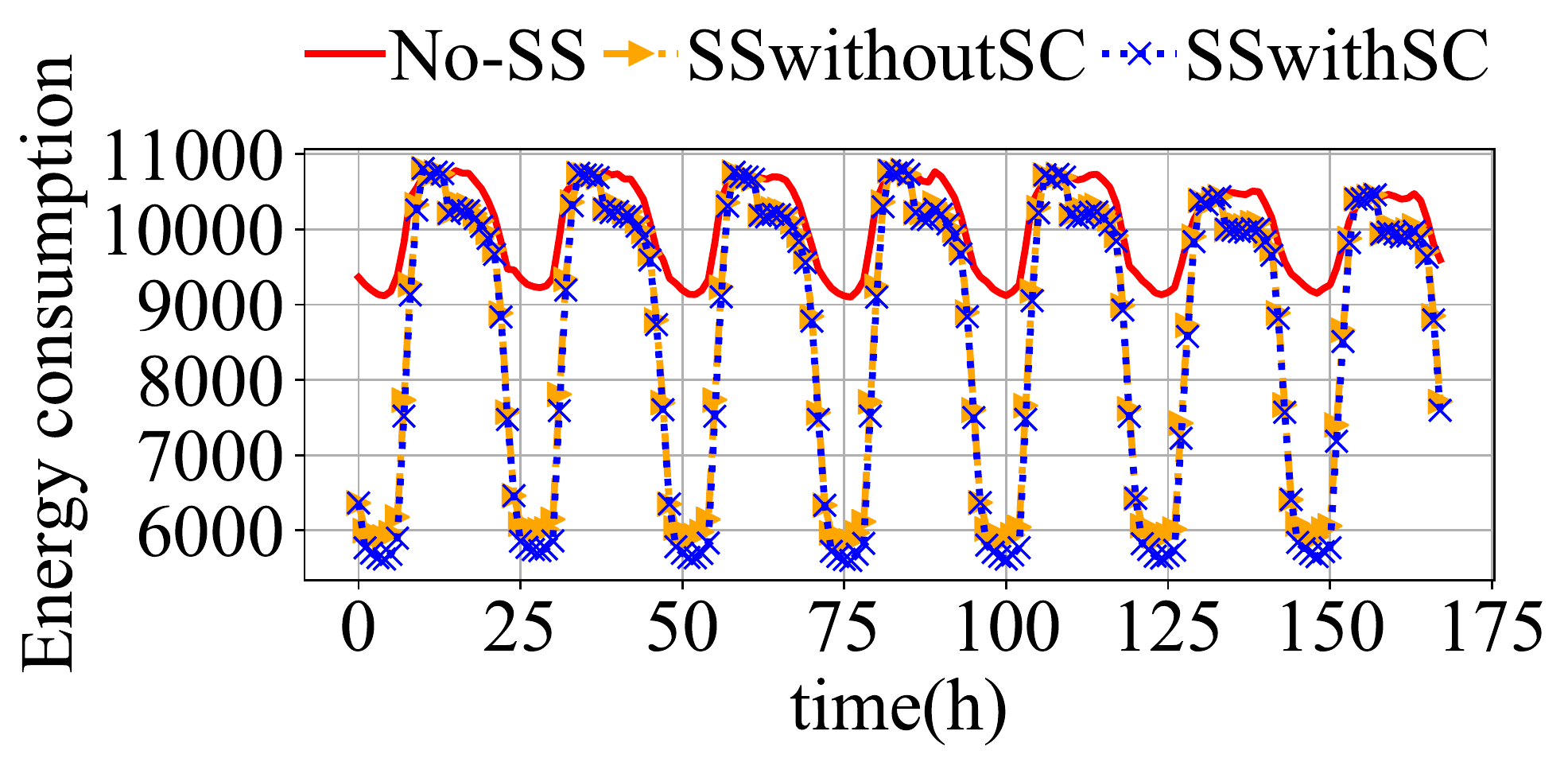}
    }
    \subfigure[Benefit of accounting for switching cost explicitly.] {
        \label{fig:draw_energy_consumption_count}
        \includegraphics[width=0.64\columnwidth]{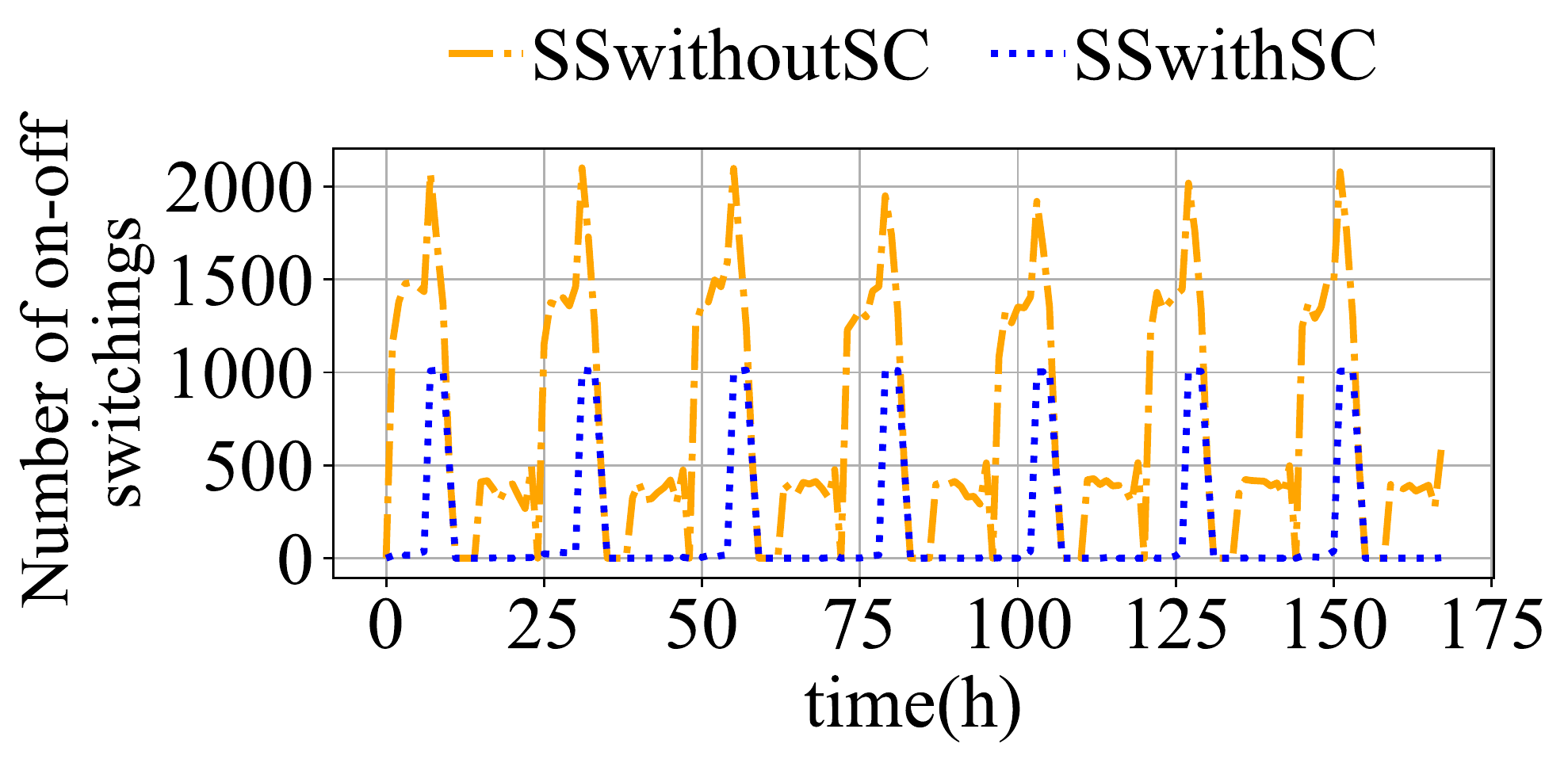}
    }
\caption{Evaluation of server scheduling.}
\label{fig:draw_energy_consumption}
\end{figure*}




We use two weeks of real traces to evaluate the workload rejection rate for different server placement policies. In Fig.~\ref{fig:diff_method_error}, we fix the total number of servers as 8000. Apparently, our approach achieves the lowest workload rejection rate. In Fig.~\ref{fig:draw_error_server_plot}, we vary the number of servers from 7000 to 12000, and study how many servers are required in order to achieve close-to-zero workload rejection rate. Under our policy, approximately 7500 servers are required. In contrast, the second best option requires about 10000 servers.



\subsection{Understand Different Design Choices of RO-RP}
\subsubsection{Evaluating different rounding schemes}\label{sec:evaluate_rounding}
To reduce the computational complexity of integer programming, we use linear programming to obtain a fractional solution first and then use a rounding scheme to get an integer solution (see Section \ref{sec:reducec_algorithmic_complexity}). Therefore, the effects of different rounding schemes should be evaluated.

We use the synthetic trace to evaluate the workload rejection rate of different rounding schemes, because our synthetic trace has a better coverage over the potential bursts than the real trace. In Fig.~\ref{fig:overload_round_error_rate_cdf}, we evaluate all the rounding schemes proposed in Section \ref{sec:reducec_algorithmic_complexity}.
Evidently, The Smallest  Resource  Pool  First Scheme has the best performance. In contrast, the Largest Resource Pool First Scheme performs the worst. 





\subsubsection{Average workload vector vs. component-wise max workload vector}
\label{sec:average-max-comparison}
In Section \ref{sec:robust_optimization}, we use the average workload vector as the representative workload vectors. Another option is to use the component-wise max workload vector. There may not be a conclusive answer that one is better than another. We compare these two options using different synthetic traces. As shown in Fig.~\ref{fig:overload_robust_balance_pool}, using average workload vector yields lower workload rejection rate for one trace, but leads to higher rejection rate for the other one.






\subsubsection{Understanding the effect of robust optimization and resource pooling} Our server placement policy utilizes both robust optimization and resource pooling to improve its robustness against workload uncertainty. To understand the contribution of each technique, we evaluate four options below:
\begin{enumerate}
    \item \textbf{Robust optimization$+$resource pooling (RO-RP):} Our server placement strategy proposed in Section \ref{sec:server_placement}.
    \item \textbf{Robust optimization only (RO-only):} Only solve (\ref{formulation:server_placement}) for a server placement solution.
    \item \textbf{Resource pooling only (RP-only):} Use only one workload vector (e.g., historical average) to solve (\ref{formulation:server_placement}) \& (\ref{formulation:server_placement_robustification}).
    \item \textbf{Not handling workload uncertainty (TwithLB):} Use only one representative workload vector to solve (\ref{formulation:server_placement}).
\end{enumerate}
From Fig.~\ref{fig:overload_error_rate_cdf_80000}, we can see that both techniques help reduce the workload rejection rate, and the performance is the best when both techniques are enabled.




\subsection{Evaluate Server Scheduling}
The objective of server scheduling is to reduce cost without affecting service quality. We compare three strategies below:
\begin{enumerate}
    \item \textbf{No server scheduling (No-SS):} This strategy turns on all the deployed edge servers at all times.
    \item \textbf{Server scheduling without accounting for the switching cost (SSwithoutSC):} Switching cost can be high. Frequently toggling a server between on/off states may reduce this server's life time.
    \item \textbf{Server scheduling that explicitly accounts for the switching cost (SSwithSC):} This is our final server scheduling strategy. Two settings for the switching cost are evaluated: 1) the switching cost is equal to running a server in idle state for 12 minutes; 2) the switching cost is equal to running a server in idle state for one hour. We also evaluate the number of on-off switchings to demonstrate the benefits of this server scheduling policy.
\end{enumerate}

\begin{figure}[!htb]
    \centering
    \includegraphics[width=0.8\linewidth]{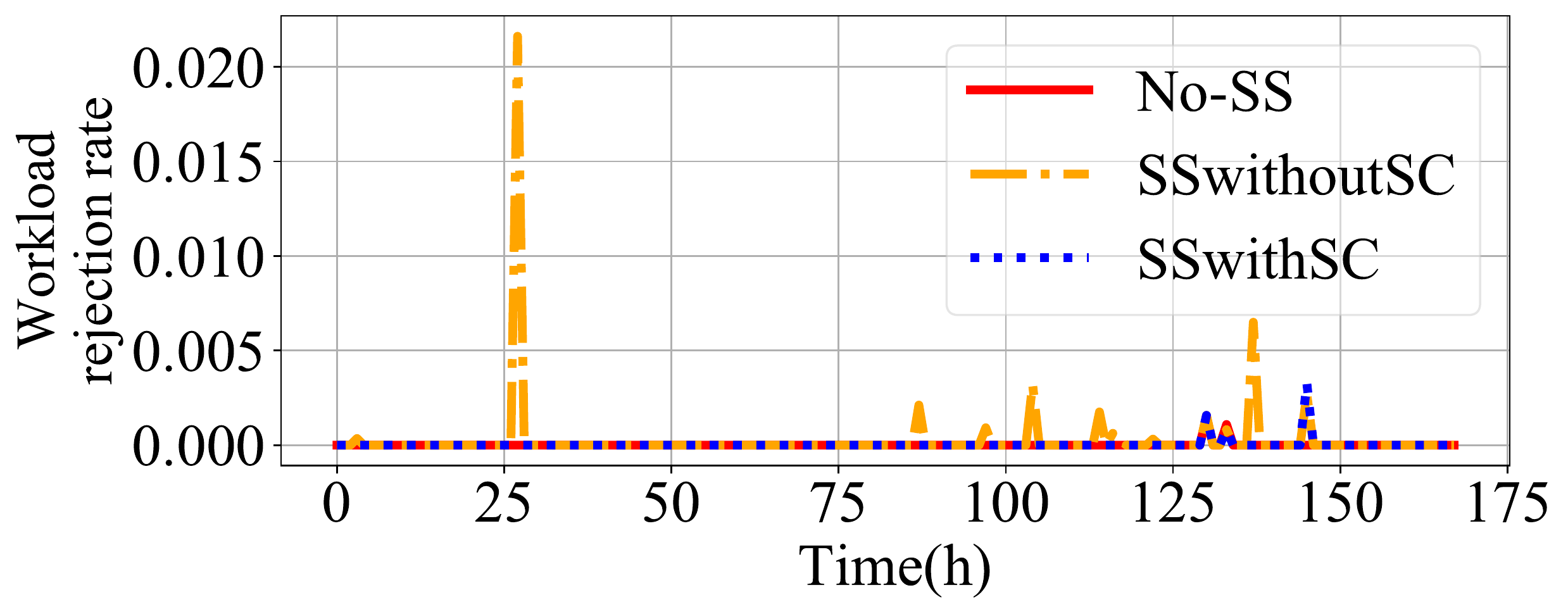}
    \caption{Workload rejection rate under different server scheduling policies.}
    \label{fig:draw_server_scheduling_final_error}
\end{figure}

We first compare the energy consumption of the three strategies in Fig.~\ref{fig:draw_energy_consumption_1}\&\ref{fig:draw_energy_consumption_02}. As expected, with server scheduling, many servers are switched off during idle hours, thus reducing the energy consumption dramatically. In total, server scheduling could save about 13\% of energy consumption. 

However, toggling servers between on/off states may incur additional cost, and thus it is better to minimize the total number of on/off switchings. As shown in Fig.~\ref{fig:draw_energy_consumption_count}, explicitly accounting for the switching cost reduces the total number of switchings by about 80\% compared to SSwithoutSC. 

Finally, we plot Fig.~\ref{fig:draw_server_scheduling_final_error} to evaluate the workload rejection rate before and after performing server scheduling. The takeaway message is that, performing server scheduling could significantly reduce energy cost, without dramatically increasing the workload rejection rate.

\section{Conclusion}
In this paper, we propose a new methodology to design server placement and server scheduling policies that are robust to workload uncertainty. This methodology utilizes robust optimization to provide guarantee for workloads that are within a predetermined uncertainty set, and performs resource pool optimization to improve service quality for out-of-bound workloads. Simulation results demonstrate the effectiveness of this methodology. From a service provider's aspect, the resulting server placement policy reduces the number of required edge servers by about 25\% compared with the state-of-the-art approach and the resulting server scheduling policy reduces the energy consumption by about 13\%.



\bibliographystyle{IEEEtran}
\bibliography{bibliography.bib}



\end{document}